\documentclass[english,aps,pra,superscriptaddress,floatfix,notitlepage,reprint,pdftex,unicode=true,colorlinks=true,citecolor=Blue,linkcolor=RubineRed,urlcolor=Blue]{revtex4-2}
\usepackage{graphicx}
\usepackage{mathrsfs}
\usepackage{bm}
\usepackage{amsmath}
\usepackage{dcolumn}
\usepackage{epstopdf}
\usepackage{dsfont}
\usepackage{amssymb}
\usepackage{tabularx}
\usepackage{array}
\usepackage{float}
\usepackage{color}
\usepackage{epstopdf}
\usepackage{mathrsfs}
 \usepackage{booktabs}
\usepackage[colorlinks, linkcolor=blue,anchorcolor=blue,citecolor=blue,urlcolor=blue]{hyperref}

 \usepackage{amsmath}

\newcommand{\ket}[1]{|#1\rangle}
\newcommand{\bra}[1]{\langle #1|}

\newcommand{\n}{\nonumber\\}

\newcommand{\abs}[1]{\lvert #1\rvert}

\newcommand{\up}{\uparrow}

\newcommand{\dw}{\downarrow}
\newcommand{\ex}[1]{\langle #1\rangle}

\begin{document}

%

\title{Quantum Charging Advantage from Multipartite Entanglement}

  \author{Hai-Long Shi}
 \email{hailong.shi@ino.cnr.it} 
\affiliation{QSTAR, INO-CNR, and LENS, Largo Enrico Fermi 2, 50125 Firenze, Italy} 
\author{Li Gan}
\affiliation{Galileo Galilei Institute for Theoretical Physics, INFN, Largo Enrico Fermi 2, 50125 Firenze, Italy}

\author{Kun Zhang}
\email{kunzhang@nwu.edu.cn}
    \affiliation{School of Physics, Northwest University, Xi'an 710127, China}
    \affiliation{Shaanxi Key Laboratory for Theoretical Physics Frontiers, Xi'an 710127, China}
    \affiliation{Peng Huanwu Center for Fundamental Theory, Xi'an 710127, China}
    
    \author{Xiao-Hui Wang}
    \email{xhwang@nwu.edu.cn}
    \affiliation{School of Physics, Northwest University, Xi'an 710127, China}
    \affiliation{Shaanxi Key Laboratory for Theoretical Physics Frontiers, Xi'an 710127, China}
    \affiliation{Peng Huanwu Center for Fundamental Theory, Xi'an 710127, China}
\author{Wen-Li Yang}
\affiliation{Institute of Modern Physics, Northwest University, Xi'an 710127, China}
\affiliation{Shaanxi Key Laboratory for Theoretical Physics Frontiers, Xi'an 710127, China}
    \affiliation{Peng Huanwu Center for Fundamental Theory, Xi'an 710127, China}

\begin{abstract}
Collective quantum batteries (QBs) demonstrate remarkable acceleration in charging dynamics compared to their individual counterparts, underscoring the pivotal contribution of quantum correlations to advanced energy storage paradigms.
A fundamental challenge lies in identifying QBs that exhibit genuine quantum advantages derived from multipartite entanglement. 
In this Letter, based on numerical and analytical evidence, we conjecture a universal bound on the charging rate for fully charging schemes, which is determined by the maximum entanglement depth arising during the charging dynamics. 
Here, the charging rate quantifies the intrinsic evolution speed of the charging process, appropriately normalized against the quantum speed limit (QSL). 
We analytically validate this conjecture in three distinct scenarios:
(i) fully charging schemes  saturating the QSL, (ii) fully parallel charging schemes, and (iii) the SU(2) fully charging schemes. 
Moreover, we establish a novel lower bound for entanglement depth detection, facilitating  numerical verification of our proposed conjecture.
By defining the genuine quantum charging advantage as the ratio between entanglement-enhanced charging rates and the maximum achievable non-entangling charging rate, we demonstrate that the charging rate constitutes a robust indicator of genuine quantum advantages.

\end{abstract}
	
\maketitle

\emph{Introduction.}--
Recent advances in quantum technologies have demonstrated the unparalleled potential of quantum correlations to outperform classical systems in computing~\cite{divincenzo1995quantum,ladd2010quantum}, communication~\cite{gisin2007quantum}, and sensing~\cite{giovannetti2011advances,RevModPhys.90.035005,RevModPhys.89.035002}.
In the field of quantum nonequilibrium thermodynamics, a novel energy storage paradigm—quantum batteries (QBs)—has emerged~\cite{PhysRevE.87.042123}, offering the prospect of harnessing multipartite entanglement among battery cells to enhance both charging speed and work extraction efficiency. 
Establishing QBs as a viable quantum technology necessitates a rigorous understanding of the critical role of entanglement in optimizing their performance~\cite{PRXQuantum.3.020101}.
Notably, entangling operations have been shown to facilitate faster work extraction~\cite{PhysRevLett.111.240401}, while extensive investigations have explored the interplay of entanglement, coherence, and quantum discord in energy storage and extraction~\cite{PhysRevLett.129.130602,PhysRevX.5.041011,PhysRevLett.122.047702,PhysRevB.104.245418,PhysRevE.105.L052101,francica2017daemonic,PhysRevE.102.042111,caravelli2021energy,PhysRevLett.125.180603,PhysRevLett.121.120602}.

A central challenge in QB research involves elucidating the role of entanglement during the charging process.
To address this, two distinct charging schemes have been proposed: the parallel charging scheme, wherein 
$N$ battery cells undergo independent charging via local unitary operations, and the collective charging scheme, which employs global unitary operations to charge all cells simultaneously.
The collective approach can achieve charging power enhancement of up to $N$-fold compared to the parallel scheme~\cite{binder2015quantacell}, a phenomenon termed quantum charging advantage. 
However, when collective charging is constrained to at most
$k$ cells, this advantage becomes bounded by $k$~\cite{PhysRevLett.118.150601,PhysRevLett.128.140501,sarkar2025hamiltonian}, demonstrating that extensive quantum advantages require fully global operations~\cite{PhysRevLett.128.140501}.
This implies a potential connection between dynamical generation of entanglement and charging performance.
Nonetheless,  
global charging dynamics do not necessarily indicate the emergence of highly entangled states. 
For instance, the Dicke QB~\cite{PhysRevLett.120.117702} achieves its charging advantage through classical effects rather than entanglement~\cite{RevModPhys.96.031001,PhysRevB.99.205437}, even when
$N$ cells interact collectively via a shared cavity. 
These findings motivate the identification of QBs that demonstrate genuine quantum advantage driven by multipartite entanglement.

Multipartite entanglement in QB can be quantified through entanglement depth~\cite{guhne2005multipartite}, which measures the maximum number of genuinely entangled battery cells. 
Quantum Fisher information (QFI) provides a useful bound for estimating entanglement depth~\cite{PhysRevA.85.022321,PhysRevLett.126.080502,PhysRevA.85.022322,imai2024metrological}. 
Using this approach, Ref.~\cite{PhysRevResearch.2.023113} demonstrated that the charging power of QBs is upper-bounded by QFI.
Nevertheless, detecting multipartite entanglement in nonequilibrium charging dynamics remains a significant challenge and is the main obstacle to identifying QBs with genuine charging advantages.
To date, the genuine quantum advantage has been rigorously demonstrated in only two many-body systems: the Sachdev-Ye-Kitaev QB~\cite{PhysRevLett.125.236402, rosa2020ultra} and the nonlinear bosonic QB~\cite{andolina2024genuine}.
It is thus crucial to develop practical tools for detecting multipartite entanglement.
Despite these advances, the quantitative relationship between genuine quantum charging advantages and multipartite entanglement remains an open question.

In this Letter, we first derive a new lower bound, Eq.~(\ref{Detect-Ent}), for estimating entanglement depth.
Compared to QFI, our bound is computationally simpler and reveals that a large weight of locally orthonormal pairs in a pure state is crucial for achieving large entanglement depth.
With this theoretical tool, we focus on fully charging schemes and propose a conjectured universal bound on the charging rate, Eq.~(\ref{conjucture-1}),  which is fundamentally constrained by the maximum entanglement depth arising during the charging dynamics. 
We introduce the charging rate to characterize the inherent evolution speed of QBs, normalized by the quantum speed limit (QSL).
We rigorously validate this conjecture in several fully charging schemes, including parallel charging, maximum-rate charging, and SU(2) charging dynamics. 
Specifically, we find that saturating the QSL in fully charging schemes requires the dynamical generation of genuine multipartite entanglement.
In parallel charging schemes, we identify the maximum achievable charging rate in non-entangling dynamics, establishing a definitive benchmark for detecting genuine quantum charging advantages.
For SU(2) dynamics, the conjectured bound can be saturated  by using charging Hamiltonian with many-body interactions.
Our findings extend previous studies on QSL~\cite{giovannetti2003role,xu2016detecting} into the regime of multipartite entanglement.

Furthermore, by defining the genuine quantum charging advantage as the ratio of the entanglement-enhanced charging rate to the maximum non-entangling rate, our conjecture implies that the charging advantage is lower-bounded by the multipartite entanglement generated during the charging dynamics. 
This establishes the charging rate as a robust indicator of genuine quantum advantage.
Our results provide new insights into the longstanding question regarding the interplay between multipartite entanglement and charging performance by identifying the minimal entanglement depth required to realize genuine quantum charging advantages in QBs, Eq.~(\ref{conjecture-2}).

\emph{Preliminaries.}--
We consider $N$-qubit QBs charged via quench dynamics, governed by the Hamiltonian
\begin{eqnarray}
H(t)=H_b+\theta(t)(H_c-H_b),
\end{eqnarray}
where $H_b\!=\!\sum_{j=1}^N h_j$ is the free Hamiltonian of  $N$ battery cells and  $H_c$ describes their many-body interactions during the charging process. 
Here, $\theta(t)$ is the Heaviside step function.
Without loss of generality, we choose $h_j\!=\!\sigma^z_j$, where $\sigma_j^\alpha$ ($\alpha\!=\!x,y,z$) are the spin-1/2 Pauli matrices.
The battery  is initially prepared in the ground state of  $H_b$, i.e., $\ket{\psi(0)}\!\equiv\! \ket{\dw}^{\otimes N }$.
The system then evolves as 
 $\ket{\psi(t)}\!=\!\exp(-iH_ct/\hbar)\ket{\psi(0)}$.
For convenience, we set $\hbar\!=\!1$.
In this Letter, we focus on the charging scheme with the maximum stored work—the fully charging scheme—where the target state after charging is the highest-energy eigenstate of $H_b$, namely $\ket{\psi(T)}\!=\!\ket{\up}^{\otimes N }$~\footnote{
The target state $\ket{\psi(T)}$ may include an arbitrary phase factor, i.e., $\ket{\psi(T)}=e^{i\phi}\ket{\up}^{\otimes N}$.
However, in many cases, computations and proofs remain valid regardless of this phase choice.
Therefore, we set it to 0 unless doing so would affect generality.
}. 
The stored work in the battery is given by  $W(T)\!\equiv\! \langle \psi(T) | H_b | \psi(T) \rangle \!-\! \langle \psi(0) | H_b | \psi(0) \rangle\! =\! 2N$, which is both maximized and fixed.
Therefore, our analysis focuses on the charging time $T$, which directly determines the charging power $W(T)/T$.

The QSL imposes a fundamental lower bound on the charging time $T$~\cite{deffner2017quantum,mandelstam1945uncertainty,margolus1998maximum,PhysRevLett.103.160502}
\begin{eqnarray}\label{QSL}
T\geq \tau_{\rm QSL}\equiv \max\left\{\frac{\pi }{2\Delta H_c}, \frac{\pi }{2\ex{H_c}}\right\},
\end{eqnarray}
where $\ex{H_c}\!\equiv\! \ex{\psi(0)|H_c|\psi(0)}$ denotes the initial interaction energy and $\Delta H_c\!\equiv\! \sqrt{\ex{H_c^2}-\ex{H_c}^2}$ represents the fluctuation in the initial interaction energy.
To compare different charging schemes on equal footing,  we introduce the charging rate $\eta$, normalized by the QSL, as follows
\begin{eqnarray}\label{charging rate}
\eta \equiv \frac{\tau_{\rm QSL}}{T}\leq 1,
\end{eqnarray}
which quantifies the intrinsic evolution speed of the charging dynamics.
Different from  Ref.~\cite{PhysRevLett.118.150601}, which separately constrains $H_c$ via energy fluctuation (C1) or initial energy (C2) bounds, we adopt the unified QSL bound, which has been proven to be tight~\cite{PhysRevLett.103.160502}.

Due to the intrinsic properties of the charging rate, different charging protocols can be classified based on the value of their charging rate:
\begin{eqnarray}\label{set}
\mathcal U(\eta)\equiv\left\{U: \ket{\up}^{\otimes N}=U(T)\ket{\dw}^{\otimes N},\  \frac{\tau_{\rm QSL}}{T}\!=\!\eta\right\},
\end{eqnarray}
which includes all fully charging schemes $U(t)\!\equiv\!\exp(-iH_c t)$ that achieve the same charging rate $\eta$.
For a given $U\!\in\!\mathcal  U(\eta)$, we define the required charging resource ${\rm Ent}[U]$ as the maximum entanglement depth of the evolved state $\ket{\psi(t)}\!=\! U(t)\ket{\psi(0)}$
\begin{eqnarray}\label{charging resource}
{\rm Ent}[U]\equiv\max_{0\leq t\leq T}{\rm Ent}[\ket{\psi(t)}],
\end{eqnarray}
where ${\rm Ent}[\ket{\psi(t)}]$ quantifies the number of genuinely entangled battery cells at time $t$~\cite{guhne2005multipartite,PhysRevLett.86.4431} (a rigorous definition will be provided in later sections).

\emph{A universal bound on quantum charging advantage.}-- 
For any fully charging dynamics $U$ with a charging rate $\eta$, the required charging resource~(\ref{charging resource}) is lower-bounded by $\left\lceil N\eta^2
\right\rceil$, i.e.,
\begin{eqnarray}\label{conjucture-1}
  {\rm Ent}[U] \geq \left\lceil N\eta^2
\right\rceil, \quad \forall U\in\mathcal U(\eta),
\end{eqnarray}
where $\lceil x\rceil$ denotes the smallest integer greater than or equal to  $x$. 

Before verifying this conjecture, we first highlight its significance.
Equation~\eqref{conjucture-1} establishes a fundamental connection between the charging rate in the fully charging scheme and the entanglement depth generated during the charging process.
Specifically, for $\eta\!=\!1$, Eq.~(\ref{conjucture-1}) leads to ${\rm Ent}[U] \!=\! N$, indicating that genuine 
$N$-particle entanglement is necessary for achieving the maximum charging rate.
Conversely, in the fully parallel charging scheme with ${\rm Ent}[U]\!=\!1$, Eq.~\eqref{conjucture-1} constrains the maximum charging rate to  $\eta_0\!\equiv\!1/\sqrt{N}$.
This observation naturally leads to the definition of the genuine charging advantage $\Gamma$, given by the ratio of the charging rate  to the maximum achievable rate under parallel charging schemes:
\begin{eqnarray}
   \Gamma \equiv \frac{\eta}{\eta_0}=\sqrt{N} \eta.
 \end{eqnarray} 
Consequently, Eq.~(\ref{conjucture-1}) can be rewritten as
\begin{eqnarray}\label{conjecture-2}
\lceil\Gamma^2\rceil\leq  {\rm Ent}[U],
\end{eqnarray}
which establishes an upper bound on the quantum charging advantage in terms of multipartite entanglement.

\emph{A new tool for estimating entanglement depth.}--
To verify the above conjecture,  we first introduce the concept of entanglement depth and derive a computationally tractable lower bound for it.
A pure state $\ket{\psi}$ of an $N$-particle system  is called $k$-producible and $h$-separable~\cite{guhne2005multipartite,szalay2019k} if it can be expressed as  $\ket{\psi}\! =\! \otimes_{\ell=1}^h \ket{\psi_\ell}$, 
where each $\ket{\psi_\ell}$ describes a state of $N_\ell \leq k$ particles, and the total particle number satisfies  $\sum_{\ell=1}^h N_\ell \!=\! N$.  
The state $\ket{\psi}$ is said to exhibit  $k$-particle entanglement if it is $k$-producible but not $(k\!-\!1)$-producible. 
The entanglement depth of such a state  $\ket{\psi}$ is defined as $k$, denoted by ${\rm Ent}[\ket{\psi}] \!=\! k$.

\emph{Theorem 1.}-- 
Let $p_0$ and $p_{\bar 0}$ be the superposition coefficients of an arbitrary locally orthonormal pair in an $N$-qubit pure state, i.e., $\ket{\psi}=p_0\ket{v_0}+p_{\bar 0}\ket{v_{\bar 0}}+\cdots$.
The entanglement depth of $\ket{\psi}$ is  then bounded from below as
\begin{eqnarray}\label{Detect-Ent}
{\rm Ent}[\ket{\psi}]\!\geq\!\left\lceil\frac{N}{\left\lfloor\log_21/|p_0p_{\bar 0}|\right\rfloor}\right\rceil.
\end{eqnarray}

We say that $\ket{v_0}\!=\!\otimes_j\ket{\nu_0}_j$ and $\ket{v_{\bar0}}\!=\!\otimes_j\ket{\nu_{\bar0}}_j$ form a local orthogonal pair if  $\ex{\nu_0|\nu_{\bar 0}}_j\!=\!0$ for  each $j$-th qubit.
A paradigmatic example illustrating Theorem 1 is the  Greenberger-Horne-Zeilinger state $\ket{{\rm GHZ}}\!=\! (\ket{0}^{\otimes N } + \ket{1}^{\otimes N })/\sqrt{2}$, which is precisely a locally orthonormal pair with coefficients $p_0 \!=\! p_{\bar 0} \!=\! 1/\sqrt{2}$ and thus exhibits $N$-particle entangled.
The central insight of Theorem 1 is that the product $|p_0 p_{\bar 0}|$, which characterizes the weight of a locally orthonormal pair, constrains the degree of separability for $\ket{\psi}$, and consequently bounds its entanglement depth.
For a detailed proof, see the Supplementary Material (SM)~\cite{SM}.

Compared to the QFI, which requires full knowledge of the quantum state, our Theorem~1 introduces a novel approach that relies only on two coefficients associated with a locally orthonormal pair. 
This significantly simplifies calculations, making our method more efficient and practical for estimating entanglement depth.
Moreover, it provides insights into how nonequilibrium dynamics can generate highly entangled states.
In essence, a substantial contribution from a locally orthonormal pair in a pure state is sufficient to ensure high entanglement.
Finally, Theorem~1 provides a numerical approach to verifying conjecture~\eqref{conjucture-1} by checking whether the maximum lower bound over time is at least $\left\lceil N\eta^2
\right\rceil$.
Examples can be found in the following sections.
Next, we provide an analytical verification of our conjecture in three typical fully charging schemes.

\emph{Maximum-rate charging.}--
Recall that the maximum charging rate $\eta\!=\!1$ corresponds to the saturation of the QSL.
The conjecture in Eq.~(\ref{conjucture-1}) implies that ${\rm Ent}[U]\!=\!N$ for $U\!\in\!\mathcal U(\eta\!=\!1)$.
To prove this, we note that  the QSL in Eq.~(\ref{QSL}) is saturated if and only if the initial state can be written as~\cite{PhysRevLett.103.160502}   
\begin{eqnarray}
\ket{\psi_{\rm QSL}(0)}\!=\!(\ket{E_0}+e^{i\phi}\ket{E_1})/\sqrt{2},
\end{eqnarray}
where $H_c\ket{E_k}\!=\!kE_1$ for $k\!=\!0,1$.
The corresponding time-evolved state is
$
\ket{\psi_{\rm QSL}(t)}\!=\!(\ket{E_0}+e^{-iE_1t}e^{i\phi}\ket{E_1})/\sqrt{2},
$
with a charging time  $T\!=\!\pi/E_1$.
Furthermore, the fully  charging condition requires that $\ket{\psi_{\rm QSL}(0)}\!=\! \ket{\dw}^{\otimes N }$  and $\ket{\psi_{\rm QSL}(T)}\!=\!  \ket{\up}^{\otimes N }$.
Consequently, the relevant low-energy eigenstates are given by $\ket{E_0}\!=\!(\ket{\dw}^{\otimes N }+\ket{\up}^{\otimes N })/\sqrt{2}$ and $\ket{E_1}\!=\!e^{-i\phi}(\ket{\dw}^{\otimes N }-\ket{\up}^{\otimes N })/\sqrt{2}$.
At half of the charging time, the evolved state simplifies to
\begin{eqnarray}
\ket{\psi_{\rm QSL}(T/2)}\!=\!\frac{1-i}{2}\ket{\dw}^{\otimes N }+\frac{1+i}{2}\ket{\up}^{\otimes N }.
\end{eqnarray}
By Theorem 1, the entanglement depth of this state is $N$, since it forms a locally orthonormal pair with equal absolute coefficients $|p_0|\!=\!|p_{\bar 0}|\!=\!1/\sqrt{2}$.

Recent work~\cite{andolina2024genuine} demonstrated that nonlinear bosonic QBs saturating the QSL exhibit a genuine quantum advantage. 
In this study, we provide a model-independent proof that $N$-particle entanglement is necessary for QSL saturation in fully charging schemes, thereby reinforcing the generality of this phenomenon.
Notably, the charging rate proposed herein serves as a diagnostic tool for identifying genuine charging advantages.
We now examine parallel (non-entangling) charging schemes to further illuminate these findings.

\emph{Parallel charging.}--
The charging Hamiltonian for the fully parallel charging scheme is given by
$H_c=\sum_{j=1}^N h_j$ where each local Hamiltonian is defined as $h_j=\alpha_j (\sigma_j^x\cos\theta+\sigma_j^y\sin\theta)/2+\alpha_j /2$ with $\alpha_j\!>\!0$.
The time-evolved state (up to an overall phase) reads
\begin{eqnarray}
\ket{\psi(t)}&=&\bigotimes_{j=1}^N \left[\cos\left(\frac{\alpha_j t}{2}\right)\ket{\dw}_j-ie^{-i\theta}\sin\left(\frac{\alpha_jt}{2}\right)\ket{\up}_j\right].\nonumber
\end{eqnarray} 
The charging time $T$ is determined as the minimum time for which $\ket{\psi(T)}=\ket{\up}^{\otimes N}$, which implies 
\begin{eqnarray}\label{parallel-T}
T=\frac{\pi(1+4 k_j)}{\alpha_j}, \ k_j\in \mathbb Z_0^+, \ \forall j=1,\ldots,N.
\end{eqnarray}%
Under these conditions, the corresponding QSL~(\ref{QSL}) reduces to
\begin{eqnarray}\label{parallel-tau}
\tau_{\rm QSL}=\frac{\pi}{\sqrt{\sum_j\alpha_j^2}}.
\end{eqnarray}
From Eq.~(\ref{parallel-T}), we obtain 
$
\alpha_j=\alpha_1(1+4k_j)/(1+4k_1).$
Substituting Eqs.~(\ref{parallel-T}) and (\ref{parallel-tau}) into the definition of charging rate~(\ref{charging rate}), we obtain 
\begin{eqnarray}\label{parallel-eta}
\eta
&=&\frac{1}{\sqrt{\sum_{j=1}^N \left(1+4k_j\right)^2} }\leq \frac{1}{\sqrt{N}},
\end{eqnarray}
since $k_j\in\mathbb Z_0^+$. 
 In particular, equality is achieved if and only if $k_j\!=\!0$, which is equivalent to all $h_j$
 being identical for all $j$.
The result~(\ref{parallel-eta}) is exactly as expected by the conjecture~(\ref{conjucture-1}): ${\rm Ent}[U]\!=\!1$ implies $\eta\leq 1/\sqrt{N}$.
In other words, the maximum charging rate in fully parallel charging schemes is \(1/\sqrt{N}\).

The physical interpretation of this result is that the maximum charging rate in fully parallel charging schemes is achieved exclusively in the homogeneous configuration, where all local charging Hamiltonians $h_j$ are identical. 
Any inhomogeneity leads to a reduction in the charging rate.
A comparison with the maximum-rate charging scheme demonstrates that the charging rate is inherently linked to the emergence of genuine multipartite entanglement. 
Therefore, the charging rate serves as an indicator of genuine charging advantages.

We further note that fully parallel charging schemes cannot achieve all values of $\eta\!<\!1/\sqrt{N}$.
For instance, as shown in Eq.~(\ref{parallel-eta}), irrational values of $\eta$ are strictly unattainable under non-entangling charging schemes.
Thus, entangling dynamics are required to realize such charging rates. 
A natural question is whether charging dynamics that generate only a low entanglement depth are sufficient to achieve these charging rates, which will be explored in future investigations. 
To further examine the limits of entanglement on charging rate, we now turn to another class of fully charging schemes with emergent SU(2) symmetry, as prior work demonstrated that fully charging phenomenon coincides with the emergence of SU(2) symmetry~\cite{PhysRevA.109.012204, PhysRevB.111.085410}.

\emph{SU(2) charging.}--
The charging Hamiltonian $H_c$ in SU(2) charging schemes can be effectively expressed as a generator $H_j$ of the spin-$d/2$ $(d\!=\!1,2,\ldots, N)$ irreducible representation of the SU(2) Lie algebra:
\begin{eqnarray}\label{SU2-H}
&&H_c=H_d\oplus H_\nu, \n 
&&H_{d}=\alpha_1 J^x+\alpha_2 J^y+\alpha_3 J^z+d|\vec\alpha|/2,
\end{eqnarray}
where $H_d$ is supported on a subspace spanned by the initial state $\ket{u_0}\!\equiv \!\ket{\dw}^{\otimes N}$,  the target state  $\ket{u_{d}}\!\equiv\!\ket{\up}^{\otimes N}$, and intermediate states $\ket{u_1}, \cdots, \ket{u_{d-1}}$.
The orthogonal component $H_\nu$ is supported on complementary subspace spanned by $\ket{w_{d+1}},\ket{w_{d+2}},\cdots, \ket{w_{2^N-1}}$. 
The explicit matrix form of $J^\alpha$ is provided in the SM~\cite{SM}.
Since $H_\nu$ does not influence fully charging dynamics, we identify the charging Hamiltonian $H_c$ with $H_d$.
Notably, for $d\!>\! 1$, the choice of intermediate state $\ket{u_{2}},\ldots, \ket{u_{d-1}}$ is arbitrary, resulting in infinitely many possible charging schemes $H_d$.
For $d\!=\!1$, the absence of intermediate states corresponds to the QSL saturation case.

As demonstrated in the SM~\cite{SM}, achieving fully charging dynamics requires $\alpha_3\!=\!0$ in Eq.~(\ref{SU2-H}).
The QSL~(\ref{QSL}) for such schemes is 
$\tau_{\rm QSL}(H_d)=\pi/\sqrt{d(\alpha_1^2+\alpha_2^2)} $.
Rewriting Eq.~(\ref{SU2-H}) as $H_d\!=\!|\vec \alpha|e^{-i\theta J_z} J_x  e^{i\theta J_z}$ with $\vec\alpha\!=\!(\alpha_1,\alpha_2,0)$ and omitting the constant term, 
the time-evolved state is obtained as 
\begin{eqnarray}\label{evolved state}
\ket{\psi(t)}\!=\!\exp(-iH_d t)\ket{\dw}^{\otimes N}=\sum_{j=0}^{d} p_j(t)\ket{u_j},
\end{eqnarray}
where 
\begin{eqnarray}\label{evolution-SU}
p_j(t)
\!=\!\sqrt{ d\choose j} 
e^{-i\theta j}
\left[-i\sin\left(\frac{|\vec \alpha|}{2}t\right)\right]^j
\left[\cos\left(\frac{|\vec \alpha|}{2}t\right)\right]^{d-j}.
\end{eqnarray}
The charging time is determined by $1\!=\!|p_{d}(T)|$, yielding $
T\!=\!\pi/|\vec\alpha|$.
Consequently, the charging rate $\eta(H_d)\!=\!\tau_{\rm QSL}(H_d)/T\!=\!1/\sqrt{d}$ depends solely on $d$, independent of the choice of intermediate states $\ket{u_{2}},\ldots, \ket{u_{d-1}}$.
For $\eta=1/\sqrt{d}$, the conjecture~(\ref{conjucture-1}) establishes a fundamental lower bound for  charging resource as ${\rm Ent}[U]\geq \left\lceil N/d
\right\rceil$.
Next, we rigorously demonstrate that this bound is not only valid but also tight within the fully SU(2) charging scheme.

\emph{Theorem 2.}--
For the fully SU(2) charging scheme, the minimal entanglement resource required to achieve a charging rate $\eta\!=\!1/\sqrt{d}$, $(d=1,2,\ldots, N)$, is $\left\lceil N/d
\right\rceil$. 

\emph{Proof.}--
For the charging Hamiltonian $H_d$, we first note from Eq.~(\ref{evolution-SU}) that
the superposition coefficients satisfy $|p_0(T/2)|\!=\!|p_{d}(T/2)|\!=\!2^{-d/2}$,  which correspond to a locally orthonormal pair.
By Theorem 1, any states generated by $H_d$ at time $T/2$ must involve at least $\left\lceil N/d
\right\rceil$  entangled particles.
To complete the proof, we explicitly construct a charging Hamiltonian, denoted as $H_d'$, for which the entanglement depth reaches its maximum at time $T/2$, attaining the exact value $\left\lceil N/d
\right\rceil$.  
We choose the intermediate states for $H_d'$ as
\begin{widetext}
\begin{eqnarray}\label{basis-state}
\ket{u_j'}=\frac{1}{\sqrt{ d\choose j} }\sum_{\mathcal P} \mathcal P(\underbrace{\ket{\up}^{\otimes k}\otimes \cdots\otimes \ket{\up}^{\otimes k} }_{j}\otimes \underbrace{\ket{\dw}^{\otimes k}\otimes \cdots\otimes \ket{\dw}^{\otimes k} \otimes\ket{\dw}^{\otimes r}}_{d-j}), 
\end{eqnarray}
where $k\!\equiv\! \left\lceil N/d
\right\rceil$, $r\!\equiv\! N\!-\!(d\!-\!1)k$, and $\mathcal P$ represents all possible permutations of the the $(d\!-\!1)$ collective $k$-particle polarized states, along with one $r$-particle polarized state.
Here, $j$ denotes the number of collective spin-up states.
Under this basis, the evolved state~(\ref{evolved state}) generated by $H_d'$ can be expressed as
\begin{eqnarray}\label{optimal evolved-state}
\ket{\psi'(t)}&\!=\!&\left[\cos\left(\frac{|\vec\alpha| t}{2}\right)\ket{\dw}^{\otimes k}\!-\!ie^{-i\theta}\sin\left(\frac{|\vec\alpha| t}{2}\right)\ket{\up}^{\otimes k}\right]^{\otimes(d-1)}\otimes \left[\cos\left(\frac{|\vec\alpha| t}{2}\right)\ket{\dw}^{\otimes r}-ie^{-i\theta}\sin\left(\frac{|\vec\alpha| t}{2}\right)\ket{\up}^{\otimes r}\right],
\end{eqnarray}
\end{widetext}
which implies that its entanglement depth is at most $k\!\equiv\! \left\lceil N/d
\right\rceil$ as it is $k$-producible.
Moreover, we have shown that the entanglement depth of all states generated by $H_d$ (including the specific case of  $H_d'$) is at least $k$.
Thus, $\ket{\psi'(t)}$ attains its maximum entanglement depth of $k$ at $T/2$, proving the Theorem 2.
Furthermore, such SU(2) charging dynamics with minimum charging resources can be realized using a Hamiltonian with $k$-body interactions
\begin{eqnarray}
H_d'\!=\!\sum_{j=1}^{d-1}\left(\bigotimes_{\ell=1}^k\sigma^x_{(j-1)k+\ell}\right)+\bigotimes_{\ell=1}^r\sigma^x_{N-\ell+1}+\frac{d}{2}, 
\end{eqnarray}
where intermediate states are given in Eq.~(\ref{basis-state}) and $\alpha_1\!=\!1$ in Eq.(\ref{SU2-H}).
These schemes are classified as hybrid charging schemes~\cite{PhysRevResearch.2.023113}.

Beyond the SU(2) charging scheme, we consider in the SM~\cite{SM} the charging Hamiltonian $ H_{c}=b_1\ket{u_1}\bra{u_0}+b_2\ket{u_2}\bra{u_1}+b_1\ket{u_3}\bra{u_2}+{\rm h.c.}$ subject to the fully charging condition.
Numerical results obtained via Theorem~1 indicate that when the Hamiltonian under consideration lacks SU(2) symmetry, the lower bound calculated from Eq.~\eqref{Detect-Ent} yields  ${\rm Ent}[U]\!=\!N$, thereby validating our conjecture.

\emph{ Conclusion.}--—
We conjecture a universal limit~(\ref{conjucture-1}) on the charging rate of QBs governed by the maximum entanglement depth generated during the fully  charging dynamics.
This conjecture is verified analytically in three paradigmatic scenarios:  maximum-rate charging, parallel charging, and SU(2) charging.
From this universal bound, we derive a constraint~(\ref{conjecture-2}) on the genuine charging advantage enabled by multipartite entanglement $\lceil\Gamma^2\rceil\leq  {\rm Ent}[U]$, emphasizing that the proposed charging rate serves as a detector of such advantage.
A related bound on the charging advantage, $\Gamma \leq \sqrt{k}$~\cite{PhysRevLett.118.150601}, has also been derived from analyses of $k$-body interaction Hamiltonians and the normalization condition of $\Delta H_c$.
Our result thus provides an alternative physical interpretation of $k$ as a measure of entanglement depth.
Moreover, since the charging rate is normalized by the QSL, our bound also establishes a direct connection between QSL and the dynamical generation of multipartite entanglement.
Methodologically, we introduce a novel approach for estimating entanglement depth, enabling both numerical verification of our conjecture and further exploration of multipartite entanglement in non-equilibrium dynamics.
Our findings provide a fundamental framework for certifying and optimizing entanglement-driven enhancements in QBs, paving the way for their development as scalable quantum technologies for energy applications.

\section*{ACKNOWLEDGMENTS}
We thank Satoya Imai, Luca Pezz\`{e}, Augusto Smerzi, and Xi-Wen Guan for helpful comments.
This work was supported by the NSFC (Grants No. 12275215, No. 12305028, No. 12247103, No. 12434006, and key Grant No. 92365202), Shaanxi Fundamental Science Research Project for Mathematics and Physics (Grant No. 22JSZ005) and the Youth Innovation Team of Shaanxi Universities.
HLS was supported  by
the Horizon Europe programme HORIZONCL4-2022-QUANTUM-02-SGA via Project No. 101113690
(PASQuanS2.1).
LG was supported by the GGI Boost Postdoctoral Fellowship (No. 25768/2023).

\bibliography{RefQB.bib}

\begin{thebibliography}{49}%
\makeatletter
\providecommand \@ifxundefined [1]{%
 \@ifx{#1\undefined}
}%
\providecommand \@ifnum [1]{%
 \ifnum #1\expandafter \@firstoftwo
 \else \expandafter \@secondoftwo
 \fi
}%
\providecommand \@ifx [1]{%
 \ifx #1\expandafter \@firstoftwo
 \else \expandafter \@secondoftwo
 \fi
}%
\providecommand \natexlab [1]{#1}%
\providecommand \enquote  [1]{``#1''}%
\providecommand \bibnamefont  [1]{#1}%
\providecommand \bibfnamefont [1]{#1}%
\providecommand \citenamefont [1]{#1}%
\providecommand \href@noop [0]{\@secondoftwo}%
\providecommand \href [0]{\begingroup \@sanitize@url \@href}%
\providecommand \@href[1]{\@@startlink{#1}\@@href}%
\providecommand \@@href[1]{\endgroup#1\@@endlink}%
\providecommand \@sanitize@url [0]{\catcode `\\12\catcode `\$12\catcode
  `\&12\catcode `\#12\catcode `\^12\catcode `\_12\catcode `\%12\relax}%
\providecommand \@@startlink[1]{}%
\providecommand \@@endlink[0]{}%
\providecommand \url  [0]{\begingroup\@sanitize@url \@url }%
\providecommand \@url [1]{\endgroup\@href {#1}{\urlprefix }}%
\providecommand \urlprefix  [0]{URL }%
\providecommand \Eprint [0]{\href }%
\providecommand \doibase [0]{https://doi.org/}%
\providecommand \selectlanguage [0]{\@gobble}%
\providecommand \bibinfo  [0]{\@secondoftwo}%
\providecommand \bibfield  [0]{\@secondoftwo}%
\providecommand \translation [1]{[#1]}%
\providecommand \BibitemOpen [0]{}%
\providecommand \bibitemStop [0]{}%
\providecommand \bibitemNoStop [0]{.\EOS\space}%
\providecommand \EOS [0]{\spacefactor3000\relax}%
\providecommand \BibitemShut  [1]{\csname bibitem#1\endcsname}%
\let\auto@bib@innerbib\@empty
\bibitem [{\citenamefont {DiVincenzo}(1995)}]{divincenzo1995quantum}%
  \BibitemOpen
  \bibfield  {author} {\bibinfo {author} {\bibfnamefont {D.~P.}\ \bibnamefont
  {DiVincenzo}},\ }\bibfield  {title} {\bibinfo {title} {Quantum computation},\
  }\href {https://doi.org/https://doi.org/10.1126/science.270.5234.255}
  {\bibfield  {journal} {\bibinfo  {journal} {Science}\ }\textbf {\bibinfo
  {volume} {270}},\ \bibinfo {pages} {255} (\bibinfo {year}
  {1995})}\BibitemShut {NoStop}%
\bibitem [{\citenamefont {Ladd}\ \emph {et~al.}(2010)\citenamefont {Ladd},
  \citenamefont {Jelezko}, \citenamefont {Laflamme}, \citenamefont {Nakamura},
  \citenamefont {Monroe},\ and\ \citenamefont {O’Brien}}]{ladd2010quantum}%
  \BibitemOpen
  \bibfield  {author} {\bibinfo {author} {\bibfnamefont {T.~D.}\ \bibnamefont
  {Ladd}}, \bibinfo {author} {\bibfnamefont {F.}~\bibnamefont {Jelezko}},
  \bibinfo {author} {\bibfnamefont {R.}~\bibnamefont {Laflamme}}, \bibinfo
  {author} {\bibfnamefont {Y.}~\bibnamefont {Nakamura}}, \bibinfo {author}
  {\bibfnamefont {C.}~\bibnamefont {Monroe}},\ and\ \bibinfo {author}
  {\bibfnamefont {J.~L.}\ \bibnamefont {O’Brien}},\ }\bibfield  {title}
  {\bibinfo {title} {Quantum computers},\ }\href
  {https://doi.org/https://doi.org/10.1038/nature08812} {\bibfield  {journal}
  {\bibinfo  {journal} {nature}\ }\textbf {\bibinfo {volume} {464}},\ \bibinfo
  {pages} {45} (\bibinfo {year} {2010})}\BibitemShut {NoStop}%
\bibitem [{\citenamefont {Gisin}\ and\ \citenamefont
  {Thew}(2007)}]{gisin2007quantum}%
  \BibitemOpen
  \bibfield  {author} {\bibinfo {author} {\bibfnamefont {N.}~\bibnamefont
  {Gisin}}\ and\ \bibinfo {author} {\bibfnamefont {R.}~\bibnamefont {Thew}},\
  }\bibfield  {title} {\bibinfo {title} {Quantum communication},\ }\href
  {https://doi.org/https://doi.org/10.1038/nphoton.2007.22} {\bibfield
  {journal} {\bibinfo  {journal} {Nature photonics}\ }\textbf {\bibinfo
  {volume} {1}},\ \bibinfo {pages} {165} (\bibinfo {year} {2007})}\BibitemShut
  {NoStop}%
\bibitem [{\citenamefont {Giovannetti}\ \emph {et~al.}(2011)\citenamefont
  {Giovannetti}, \citenamefont {Lloyd},\ and\ \citenamefont
  {Maccone}}]{giovannetti2011advances}%
  \BibitemOpen
  \bibfield  {author} {\bibinfo {author} {\bibfnamefont {V.}~\bibnamefont
  {Giovannetti}}, \bibinfo {author} {\bibfnamefont {S.}~\bibnamefont {Lloyd}},\
  and\ \bibinfo {author} {\bibfnamefont {L.}~\bibnamefont {Maccone}},\
  }\bibfield  {title} {\bibinfo {title} {Advances in quantum metrology},\
  }\href {https://doi.org/https://doi.org/10.1038/nphoton.2011.35} {\bibfield
  {journal} {\bibinfo  {journal} {Nature photonics}\ }\textbf {\bibinfo
  {volume} {5}},\ \bibinfo {pages} {222} (\bibinfo {year} {2011})}\BibitemShut
  {NoStop}%
\bibitem [{\citenamefont {Pezz\`e}\ \emph {et~al.}(2018)\citenamefont
  {Pezz\`e}, \citenamefont {Smerzi}, \citenamefont {Oberthaler}, \citenamefont
  {Schmied},\ and\ \citenamefont {Treutlein}}]{RevModPhys.90.035005}%
  \BibitemOpen
  \bibfield  {author} {\bibinfo {author} {\bibfnamefont {L.}~\bibnamefont
  {Pezz\`e}}, \bibinfo {author} {\bibfnamefont {A.}~\bibnamefont {Smerzi}},
  \bibinfo {author} {\bibfnamefont {M.~K.}\ \bibnamefont {Oberthaler}},
  \bibinfo {author} {\bibfnamefont {R.}~\bibnamefont {Schmied}},\ and\ \bibinfo
  {author} {\bibfnamefont {P.}~\bibnamefont {Treutlein}},\ }\bibfield  {title}
  {\bibinfo {title} {Quantum metrology with nonclassical states of atomic
  ensembles},\ }\href {https://doi.org/10.1103/RevModPhys.90.035005} {\bibfield
   {journal} {\bibinfo  {journal} {Rev. Mod. Phys.}\ }\textbf {\bibinfo
  {volume} {90}},\ \bibinfo {pages} {035005} (\bibinfo {year}
  {2018})}\BibitemShut {NoStop}%
\bibitem [{\citenamefont {Degen}\ \emph {et~al.}(2017)\citenamefont {Degen},
  \citenamefont {Reinhard},\ and\ \citenamefont
  {Cappellaro}}]{RevModPhys.89.035002}%
  \BibitemOpen
  \bibfield  {author} {\bibinfo {author} {\bibfnamefont {C.~L.}\ \bibnamefont
  {Degen}}, \bibinfo {author} {\bibfnamefont {F.}~\bibnamefont {Reinhard}},\
  and\ \bibinfo {author} {\bibfnamefont {P.}~\bibnamefont {Cappellaro}},\
  }\bibfield  {title} {\bibinfo {title} {Quantum sensing},\ }\href
  {https://doi.org/10.1103/RevModPhys.89.035002} {\bibfield  {journal}
  {\bibinfo  {journal} {Rev. Mod. Phys.}\ }\textbf {\bibinfo {volume} {89}},\
  \bibinfo {pages} {035002} (\bibinfo {year} {2017})}\BibitemShut {NoStop}%
\bibitem [{\citenamefont {Alicki}\ and\ \citenamefont
  {Fannes}(2013)}]{PhysRevE.87.042123}%
  \BibitemOpen
  \bibfield  {author} {\bibinfo {author} {\bibfnamefont {R.}~\bibnamefont
  {Alicki}}\ and\ \bibinfo {author} {\bibfnamefont {M.}~\bibnamefont
  {Fannes}},\ }\bibfield  {title} {\bibinfo {title} {Entanglement boost for
  extractable work from ensembles of quantum batteries},\ }\href
  {https://doi.org/10.1103/PhysRevE.87.042123} {\bibfield  {journal} {\bibinfo
  {journal} {Phys. Rev. E}\ }\textbf {\bibinfo {volume} {87}},\ \bibinfo
  {pages} {042123} (\bibinfo {year} {2013})}\BibitemShut {NoStop}%
\bibitem [{\citenamefont {Auff\`eves}(2022)}]{PRXQuantum.3.020101}%
  \BibitemOpen
  \bibfield  {author} {\bibinfo {author} {\bibfnamefont {A.}~\bibnamefont
  {Auff\`eves}},\ }\bibfield  {title} {\bibinfo {title} {Quantum technologies
  need a quantum energy initiative},\ }\href
  {https://doi.org/10.1103/PRXQuantum.3.020101} {\bibfield  {journal} {\bibinfo
   {journal} {PRX Quantum}\ }\textbf {\bibinfo {volume} {3}},\ \bibinfo {pages}
  {020101} (\bibinfo {year} {2022})}\BibitemShut {NoStop}%
\bibitem [{\citenamefont {Hovhannisyan}\ \emph {et~al.}(2013)\citenamefont
  {Hovhannisyan}, \citenamefont {Perarnau-Llobet}, \citenamefont {Huber},\ and\
  \citenamefont {Ac\'{\i}n}}]{PhysRevLett.111.240401}%
  \BibitemOpen
  \bibfield  {author} {\bibinfo {author} {\bibfnamefont {K.~V.}\ \bibnamefont
  {Hovhannisyan}}, \bibinfo {author} {\bibfnamefont {M.}~\bibnamefont
  {Perarnau-Llobet}}, \bibinfo {author} {\bibfnamefont {M.}~\bibnamefont
  {Huber}},\ and\ \bibinfo {author} {\bibfnamefont {A.}~\bibnamefont
  {Ac\'{\i}n}},\ }\bibfield  {title} {\bibinfo {title} {Entanglement generation
  is not necessary for optimal work extraction},\ }\href
  {https://doi.org/10.1103/PhysRevLett.111.240401} {\bibfield  {journal}
  {\bibinfo  {journal} {Phys. Rev. Lett.}\ }\textbf {\bibinfo {volume} {111}},\
  \bibinfo {pages} {240401} (\bibinfo {year} {2013})}\BibitemShut {NoStop}%
\bibitem [{\citenamefont {Shi}\ \emph {et~al.}(2022)\citenamefont {Shi},
  \citenamefont {Ding}, \citenamefont {Wan}, \citenamefont {Wang},\ and\
  \citenamefont {Yang}}]{PhysRevLett.129.130602}%
  \BibitemOpen
  \bibfield  {author} {\bibinfo {author} {\bibfnamefont {H.-L.}\ \bibnamefont
  {Shi}}, \bibinfo {author} {\bibfnamefont {S.}~\bibnamefont {Ding}}, \bibinfo
  {author} {\bibfnamefont {Q.-K.}\ \bibnamefont {Wan}}, \bibinfo {author}
  {\bibfnamefont {X.-H.}\ \bibnamefont {Wang}},\ and\ \bibinfo {author}
  {\bibfnamefont {W.-L.}\ \bibnamefont {Yang}},\ }\bibfield  {title} {\bibinfo
  {title} {Entanglement, coherence, and extractable work in quantum
  batteries},\ }\href {https://doi.org/10.1103/PhysRevLett.129.130602}
  {\bibfield  {journal} {\bibinfo  {journal} {Phys. Rev. Lett.}\ }\textbf
  {\bibinfo {volume} {129}},\ \bibinfo {pages} {130602} (\bibinfo {year}
  {2022})}\BibitemShut {NoStop}%
\bibitem [{\citenamefont {Perarnau-Llobet}\ \emph {et~al.}(2015)\citenamefont
  {Perarnau-Llobet}, \citenamefont {Hovhannisyan}, \citenamefont {Huber},
  \citenamefont {Skrzypczyk}, \citenamefont {Brunner},\ and\ \citenamefont
  {Ac\'{\i}n}}]{PhysRevX.5.041011}%
  \BibitemOpen
  \bibfield  {author} {\bibinfo {author} {\bibfnamefont {M.}~\bibnamefont
  {Perarnau-Llobet}}, \bibinfo {author} {\bibfnamefont {K.~V.}\ \bibnamefont
  {Hovhannisyan}}, \bibinfo {author} {\bibfnamefont {M.}~\bibnamefont {Huber}},
  \bibinfo {author} {\bibfnamefont {P.}~\bibnamefont {Skrzypczyk}}, \bibinfo
  {author} {\bibfnamefont {N.}~\bibnamefont {Brunner}},\ and\ \bibinfo {author}
  {\bibfnamefont {A.}~\bibnamefont {Ac\'{\i}n}},\ }\bibfield  {title} {\bibinfo
  {title} {Extractable work from correlations},\ }\href
  {https://doi.org/10.1103/PhysRevX.5.041011} {\bibfield  {journal} {\bibinfo
  {journal} {Phys. Rev. X}\ }\textbf {\bibinfo {volume} {5}},\ \bibinfo {pages}
  {041011} (\bibinfo {year} {2015})}\BibitemShut {NoStop}%
\bibitem [{\citenamefont {Andolina}\ \emph
  {et~al.}(2019{\natexlab{a}})\citenamefont {Andolina}, \citenamefont {Keck},
  \citenamefont {Mari}, \citenamefont {Campisi}, \citenamefont {Giovannetti},\
  and\ \citenamefont {Polini}}]{PhysRevLett.122.047702}%
  \BibitemOpen
  \bibfield  {author} {\bibinfo {author} {\bibfnamefont {G.~M.}\ \bibnamefont
  {Andolina}}, \bibinfo {author} {\bibfnamefont {M.}~\bibnamefont {Keck}},
  \bibinfo {author} {\bibfnamefont {A.}~\bibnamefont {Mari}}, \bibinfo {author}
  {\bibfnamefont {M.}~\bibnamefont {Campisi}}, \bibinfo {author} {\bibfnamefont
  {V.}~\bibnamefont {Giovannetti}},\ and\ \bibinfo {author} {\bibfnamefont
  {M.}~\bibnamefont {Polini}},\ }\bibfield  {title} {\bibinfo {title}
  {Extractable work, the role of correlations, and asymptotic freedom in
  quantum batteries},\ }\href {https://doi.org/10.1103/PhysRevLett.122.047702}
  {\bibfield  {journal} {\bibinfo  {journal} {Phys. Rev. Lett.}\ }\textbf
  {\bibinfo {volume} {122}},\ \bibinfo {pages} {047702} (\bibinfo {year}
  {2019}{\natexlab{a}})}\BibitemShut {NoStop}%
\bibitem [{\citenamefont {Liu}\ \emph {et~al.}(2021)\citenamefont {Liu},
  \citenamefont {Shi}, \citenamefont {Shi}, \citenamefont {Wang},\ and\
  \citenamefont {Yang}}]{PhysRevB.104.245418}%
  \BibitemOpen
  \bibfield  {author} {\bibinfo {author} {\bibfnamefont {J.-X.}\ \bibnamefont
  {Liu}}, \bibinfo {author} {\bibfnamefont {H.-L.}\ \bibnamefont {Shi}},
  \bibinfo {author} {\bibfnamefont {Y.-H.}\ \bibnamefont {Shi}}, \bibinfo
  {author} {\bibfnamefont {X.-H.}\ \bibnamefont {Wang}},\ and\ \bibinfo
  {author} {\bibfnamefont {W.-L.}\ \bibnamefont {Yang}},\ }\bibfield  {title}
  {\bibinfo {title} {Entanglement and work extraction in the central-spin
  quantum battery},\ }\href {https://doi.org/10.1103/PhysRevB.104.245418}
  {\bibfield  {journal} {\bibinfo  {journal} {Phys. Rev. B}\ }\textbf {\bibinfo
  {volume} {104}},\ \bibinfo {pages} {245418} (\bibinfo {year}
  {2021})}\BibitemShut {NoStop}%
\bibitem [{\citenamefont {Francica}(2022)}]{PhysRevE.105.L052101}%
  \BibitemOpen
  \bibfield  {author} {\bibinfo {author} {\bibfnamefont {G.}~\bibnamefont
  {Francica}},\ }\bibfield  {title} {\bibinfo {title} {Quantum correlations and
  ergotropy},\ }\href {https://doi.org/10.1103/PhysRevE.105.L052101} {\bibfield
   {journal} {\bibinfo  {journal} {Phys. Rev. E}\ }\textbf {\bibinfo {volume}
  {105}},\ \bibinfo {pages} {L052101} (\bibinfo {year} {2022})}\BibitemShut
  {NoStop}%
\bibitem [{\citenamefont {Francica}\ \emph {et~al.}(2017)\citenamefont
  {Francica}, \citenamefont {Goold}, \citenamefont {Plastina},\ and\
  \citenamefont {Paternostro}}]{francica2017daemonic}%
  \BibitemOpen
  \bibfield  {author} {\bibinfo {author} {\bibfnamefont {G.}~\bibnamefont
  {Francica}}, \bibinfo {author} {\bibfnamefont {J.}~\bibnamefont {Goold}},
  \bibinfo {author} {\bibfnamefont {F.}~\bibnamefont {Plastina}},\ and\
  \bibinfo {author} {\bibfnamefont {M.}~\bibnamefont {Paternostro}},\
  }\bibfield  {title} {\bibinfo {title} {Daemonic ergotropy: Enhanced work
  extraction from quantum correlations},\ }\href@noop {} {\bibfield  {journal}
  {\bibinfo  {journal} {npj Quantum Information}\ }\textbf {\bibinfo {volume}
  {3}},\ \bibinfo {pages} {12} (\bibinfo {year} {2017})}\BibitemShut {NoStop}%
\bibitem [{\citenamefont {\c{C}akmak}(2020)}]{PhysRevE.102.042111}%
  \BibitemOpen
  \bibfield  {author} {\bibinfo {author} {\bibfnamefont {B.}~\bibnamefont
  {\c{C}akmak}},\ }\bibfield  {title} {\bibinfo {title} {Ergotropy from
  coherences in an open quantum system},\ }\href
  {https://doi.org/10.1103/PhysRevE.102.042111} {\bibfield  {journal} {\bibinfo
   {journal} {Phys. Rev. E}\ }\textbf {\bibinfo {volume} {102}},\ \bibinfo
  {pages} {042111} (\bibinfo {year} {2020})}\BibitemShut {NoStop}%
\bibitem [{\citenamefont {Caravelli}\ \emph {et~al.}(2021)\citenamefont
  {Caravelli}, \citenamefont {Yan}, \citenamefont {Garc{\'\i}a-Pintos},\ and\
  \citenamefont {Hamma}}]{caravelli2021energy}%
  \BibitemOpen
  \bibfield  {author} {\bibinfo {author} {\bibfnamefont {F.}~\bibnamefont
  {Caravelli}}, \bibinfo {author} {\bibfnamefont {B.}~\bibnamefont {Yan}},
  \bibinfo {author} {\bibfnamefont {L.~P.}\ \bibnamefont
  {Garc{\'\i}a-Pintos}},\ and\ \bibinfo {author} {\bibfnamefont
  {A.}~\bibnamefont {Hamma}},\ }\bibfield  {title} {\bibinfo {title} {Energy
  storage and coherence in closed and open quantum batteries},\ }\href@noop {}
  {\bibfield  {journal} {\bibinfo  {journal} {Quantum}\ }\textbf {\bibinfo
  {volume} {5}},\ \bibinfo {pages} {505} (\bibinfo {year} {2021})}\BibitemShut
  {NoStop}%
\bibitem [{\citenamefont {Francica}\ \emph {et~al.}(2020)\citenamefont
  {Francica}, \citenamefont {Binder}, \citenamefont {Guarnieri}, \citenamefont
  {Mitchison}, \citenamefont {Goold},\ and\ \citenamefont
  {Plastina}}]{PhysRevLett.125.180603}%
  \BibitemOpen
  \bibfield  {author} {\bibinfo {author} {\bibfnamefont {G.}~\bibnamefont
  {Francica}}, \bibinfo {author} {\bibfnamefont {F.~C.}\ \bibnamefont
  {Binder}}, \bibinfo {author} {\bibfnamefont {G.}~\bibnamefont {Guarnieri}},
  \bibinfo {author} {\bibfnamefont {M.~T.}\ \bibnamefont {Mitchison}}, \bibinfo
  {author} {\bibfnamefont {J.}~\bibnamefont {Goold}},\ and\ \bibinfo {author}
  {\bibfnamefont {F.}~\bibnamefont {Plastina}},\ }\bibfield  {title} {\bibinfo
  {title} {Quantum coherence and ergotropy},\ }\href
  {https://doi.org/10.1103/PhysRevLett.125.180603} {\bibfield  {journal}
  {\bibinfo  {journal} {Phys. Rev. Lett.}\ }\textbf {\bibinfo {volume} {125}},\
  \bibinfo {pages} {180603} (\bibinfo {year} {2020})}\BibitemShut {NoStop}%
\bibitem [{\citenamefont {Manzano}\ \emph {et~al.}(2018)\citenamefont
  {Manzano}, \citenamefont {Plastina},\ and\ \citenamefont
  {Zambrini}}]{PhysRevLett.121.120602}%
  \BibitemOpen
  \bibfield  {author} {\bibinfo {author} {\bibfnamefont {G.}~\bibnamefont
  {Manzano}}, \bibinfo {author} {\bibfnamefont {F.}~\bibnamefont {Plastina}},\
  and\ \bibinfo {author} {\bibfnamefont {R.}~\bibnamefont {Zambrini}},\
  }\bibfield  {title} {\bibinfo {title} {Optimal work extraction and
  thermodynamics of quantum measurements and correlations},\ }\href
  {https://doi.org/10.1103/PhysRevLett.121.120602} {\bibfield  {journal}
  {\bibinfo  {journal} {Phys. Rev. Lett.}\ }\textbf {\bibinfo {volume} {121}},\
  \bibinfo {pages} {120602} (\bibinfo {year} {2018})}\BibitemShut {NoStop}%
\bibitem [{\citenamefont {Binder}\ \emph {et~al.}(2015)\citenamefont {Binder},
  \citenamefont {Vinjanampathy}, \citenamefont {Modi},\ and\ \citenamefont
  {Goold}}]{binder2015quantacell}%
  \BibitemOpen
  \bibfield  {author} {\bibinfo {author} {\bibfnamefont {F.~C.}\ \bibnamefont
  {Binder}}, \bibinfo {author} {\bibfnamefont {S.}~\bibnamefont
  {Vinjanampathy}}, \bibinfo {author} {\bibfnamefont {K.}~\bibnamefont
  {Modi}},\ and\ \bibinfo {author} {\bibfnamefont {J.}~\bibnamefont {Goold}},\
  }\bibfield  {title} {\bibinfo {title} {Quantacell: powerful charging of
  quantum batteries},\ }\href {https://doi.org/10.1088/1367-2630/17/7/075015}
  {\bibfield  {journal} {\bibinfo  {journal} {New Journal of Physics}\ }\textbf
  {\bibinfo {volume} {17}},\ \bibinfo {pages} {075015} (\bibinfo {year}
  {2015})}\BibitemShut {NoStop}%
\bibitem [{\citenamefont {Campaioli}\ \emph {et~al.}(2017)\citenamefont
  {Campaioli}, \citenamefont {Pollock}, \citenamefont {Binder}, \citenamefont
  {C\'eleri}, \citenamefont {Goold}, \citenamefont {Vinjanampathy},\ and\
  \citenamefont {Modi}}]{PhysRevLett.118.150601}%
  \BibitemOpen
  \bibfield  {author} {\bibinfo {author} {\bibfnamefont {F.}~\bibnamefont
  {Campaioli}}, \bibinfo {author} {\bibfnamefont {F.~A.}\ \bibnamefont
  {Pollock}}, \bibinfo {author} {\bibfnamefont {F.~C.}\ \bibnamefont {Binder}},
  \bibinfo {author} {\bibfnamefont {L.}~\bibnamefont {C\'eleri}}, \bibinfo
  {author} {\bibfnamefont {J.}~\bibnamefont {Goold}}, \bibinfo {author}
  {\bibfnamefont {S.}~\bibnamefont {Vinjanampathy}},\ and\ \bibinfo {author}
  {\bibfnamefont {K.}~\bibnamefont {Modi}},\ }\bibfield  {title} {\bibinfo
  {title} {Enhancing the charging power of quantum batteries},\ }\href
  {https://doi.org/10.1103/PhysRevLett.118.150601} {\bibfield  {journal}
  {\bibinfo  {journal} {Phys. Rev. Lett.}\ }\textbf {\bibinfo {volume} {118}},\
  \bibinfo {pages} {150601} (\bibinfo {year} {2017})}\BibitemShut {NoStop}%
\bibitem [{\citenamefont {Gyhm}\ \emph {et~al.}(2022)\citenamefont {Gyhm},
  \citenamefont {\ifmmode~\check{S}\else \v{S}\fi{}afr\'anek},\ and\
  \citenamefont {Rosa}}]{PhysRevLett.128.140501}%
  \BibitemOpen
  \bibfield  {author} {\bibinfo {author} {\bibfnamefont {J.-Y.}\ \bibnamefont
  {Gyhm}}, \bibinfo {author} {\bibfnamefont {D.}~\bibnamefont
  {\ifmmode~\check{S}\else \v{S}\fi{}afr\'anek}},\ and\ \bibinfo {author}
  {\bibfnamefont {D.}~\bibnamefont {Rosa}},\ }\bibfield  {title} {\bibinfo
  {title} {Quantum charging advantage cannot be extensive without global
  operations},\ }\href {https://doi.org/10.1103/PhysRevLett.128.140501}
  {\bibfield  {journal} {\bibinfo  {journal} {Phys. Rev. Lett.}\ }\textbf
  {\bibinfo {volume} {128}},\ \bibinfo {pages} {140501} (\bibinfo {year}
  {2022})}\BibitemShut {NoStop}%
\bibitem [{\citenamefont {Sarkar}\ and\ \citenamefont
  {Ghosh}(2025)}]{sarkar2025hamiltonian}%
  \BibitemOpen
  \bibfield  {author} {\bibinfo {author} {\bibfnamefont {A.}~\bibnamefont
  {Sarkar}}\ and\ \bibinfo {author} {\bibfnamefont {S.}~\bibnamefont {Ghosh}},\
  }\bibfield  {title} {\bibinfo {title} {Hamiltonian $ k $-locality is the key
  resource for powerful quantum battery charging},\ }\href@noop {} {\bibfield
  {journal} {\bibinfo  {journal} {arXiv preprint arXiv:2501.12000}\ } (\bibinfo
  {year} {2025})}\BibitemShut {NoStop}%
\bibitem [{\citenamefont {Ferraro}\ \emph {et~al.}(2018)\citenamefont
  {Ferraro}, \citenamefont {Campisi}, \citenamefont {Andolina}, \citenamefont
  {Pellegrini},\ and\ \citenamefont {Polini}}]{PhysRevLett.120.117702}%
  \BibitemOpen
  \bibfield  {author} {\bibinfo {author} {\bibfnamefont {D.}~\bibnamefont
  {Ferraro}}, \bibinfo {author} {\bibfnamefont {M.}~\bibnamefont {Campisi}},
  \bibinfo {author} {\bibfnamefont {G.~M.}\ \bibnamefont {Andolina}}, \bibinfo
  {author} {\bibfnamefont {V.}~\bibnamefont {Pellegrini}},\ and\ \bibinfo
  {author} {\bibfnamefont {M.}~\bibnamefont {Polini}},\ }\bibfield  {title}
  {\bibinfo {title} {High-power collective charging of a solid-state quantum
  battery},\ }\href {https://doi.org/10.1103/PhysRevLett.120.117702} {\bibfield
   {journal} {\bibinfo  {journal} {Phys. Rev. Lett.}\ }\textbf {\bibinfo
  {volume} {120}},\ \bibinfo {pages} {117702} (\bibinfo {year}
  {2018})}\BibitemShut {NoStop}%
\bibitem [{\citenamefont {Campaioli}\ \emph {et~al.}(2024)\citenamefont
  {Campaioli}, \citenamefont {Gherardini}, \citenamefont {Quach}, \citenamefont
  {Polini},\ and\ \citenamefont {Andolina}}]{RevModPhys.96.031001}%
  \BibitemOpen
  \bibfield  {author} {\bibinfo {author} {\bibfnamefont {F.}~\bibnamefont
  {Campaioli}}, \bibinfo {author} {\bibfnamefont {S.}~\bibnamefont
  {Gherardini}}, \bibinfo {author} {\bibfnamefont {J.~Q.}\ \bibnamefont
  {Quach}}, \bibinfo {author} {\bibfnamefont {M.}~\bibnamefont {Polini}},\ and\
  \bibinfo {author} {\bibfnamefont {G.~M.}\ \bibnamefont {Andolina}},\
  }\bibfield  {title} {\bibinfo {title} {Colloquium: Quantum batteries},\
  }\href {https://doi.org/10.1103/RevModPhys.96.031001} {\bibfield  {journal}
  {\bibinfo  {journal} {Rev. Mod. Phys.}\ }\textbf {\bibinfo {volume} {96}},\
  \bibinfo {pages} {031001} (\bibinfo {year} {2024})}\BibitemShut {NoStop}%
\bibitem [{\citenamefont {Andolina}\ \emph
  {et~al.}(2019{\natexlab{b}})\citenamefont {Andolina}, \citenamefont {Keck},
  \citenamefont {Mari}, \citenamefont {Giovannetti},\ and\ \citenamefont
  {Polini}}]{PhysRevB.99.205437}%
  \BibitemOpen
  \bibfield  {author} {\bibinfo {author} {\bibfnamefont {G.~M.}\ \bibnamefont
  {Andolina}}, \bibinfo {author} {\bibfnamefont {M.}~\bibnamefont {Keck}},
  \bibinfo {author} {\bibfnamefont {A.}~\bibnamefont {Mari}}, \bibinfo {author}
  {\bibfnamefont {V.}~\bibnamefont {Giovannetti}},\ and\ \bibinfo {author}
  {\bibfnamefont {M.}~\bibnamefont {Polini}},\ }\bibfield  {title} {\bibinfo
  {title} {Quantum versus classical many-body batteries},\ }\href
  {https://doi.org/10.1103/PhysRevB.99.205437} {\bibfield  {journal} {\bibinfo
  {journal} {Phys. Rev. B}\ }\textbf {\bibinfo {volume} {99}},\ \bibinfo
  {pages} {205437} (\bibinfo {year} {2019}{\natexlab{b}})}\BibitemShut
  {NoStop}%
\bibitem [{\citenamefont {G{\"u}hne}\ \emph {et~al.}(2005)\citenamefont
  {G{\"u}hne}, \citenamefont {T{\'o}th},\ and\ \citenamefont
  {Briegel}}]{guhne2005multipartite}%
  \BibitemOpen
  \bibfield  {author} {\bibinfo {author} {\bibfnamefont {O.}~\bibnamefont
  {G{\"u}hne}}, \bibinfo {author} {\bibfnamefont {G.}~\bibnamefont
  {T{\'o}th}},\ and\ \bibinfo {author} {\bibfnamefont {H.~J.}\ \bibnamefont
  {Briegel}},\ }\bibfield  {title} {\bibinfo {title} {Multipartite entanglement
  in spin chains},\ }\href {https://doi.org/10.1088/1367-2630/7/1/229}
  {\bibfield  {journal} {\bibinfo  {journal} {New Journal of Physics}\ }\textbf
  {\bibinfo {volume} {7}},\ \bibinfo {pages} {229} (\bibinfo {year}
  {2005})}\BibitemShut {NoStop}%
\bibitem [{\citenamefont {Hyllus}\ \emph {et~al.}(2012)\citenamefont {Hyllus},
  \citenamefont {Laskowski}, \citenamefont {Krischek}, \citenamefont
  {Schwemmer}, \citenamefont {Wieczorek}, \citenamefont {Weinfurter},
  \citenamefont {Pezz\'e},\ and\ \citenamefont {Smerzi}}]{PhysRevA.85.022321}%
  \BibitemOpen
  \bibfield  {author} {\bibinfo {author} {\bibfnamefont {P.}~\bibnamefont
  {Hyllus}}, \bibinfo {author} {\bibfnamefont {W.}~\bibnamefont {Laskowski}},
  \bibinfo {author} {\bibfnamefont {R.}~\bibnamefont {Krischek}}, \bibinfo
  {author} {\bibfnamefont {C.}~\bibnamefont {Schwemmer}}, \bibinfo {author}
  {\bibfnamefont {W.}~\bibnamefont {Wieczorek}}, \bibinfo {author}
  {\bibfnamefont {H.}~\bibnamefont {Weinfurter}}, \bibinfo {author}
  {\bibfnamefont {L.}~\bibnamefont {Pezz\'e}},\ and\ \bibinfo {author}
  {\bibfnamefont {A.}~\bibnamefont {Smerzi}},\ }\bibfield  {title} {\bibinfo
  {title} {Fisher information and multiparticle entanglement},\ }\href
  {https://doi.org/10.1103/PhysRevA.85.022321} {\bibfield  {journal} {\bibinfo
  {journal} {Phys. Rev. A}\ }\textbf {\bibinfo {volume} {85}},\ \bibinfo
  {pages} {022321} (\bibinfo {year} {2012})}\BibitemShut {NoStop}%
\bibitem [{\citenamefont {Ren}\ \emph {et~al.}(2021)\citenamefont {Ren},
  \citenamefont {Li}, \citenamefont {Smerzi},\ and\ \citenamefont
  {Gessner}}]{PhysRevLett.126.080502}%
  \BibitemOpen
  \bibfield  {author} {\bibinfo {author} {\bibfnamefont {Z.}~\bibnamefont
  {Ren}}, \bibinfo {author} {\bibfnamefont {W.}~\bibnamefont {Li}}, \bibinfo
  {author} {\bibfnamefont {A.}~\bibnamefont {Smerzi}},\ and\ \bibinfo {author}
  {\bibfnamefont {M.}~\bibnamefont {Gessner}},\ }\bibfield  {title} {\bibinfo
  {title} {Metrological detection of multipartite entanglement from young
  diagrams},\ }\href {https://doi.org/10.1103/PhysRevLett.126.080502}
  {\bibfield  {journal} {\bibinfo  {journal} {Phys. Rev. Lett.}\ }\textbf
  {\bibinfo {volume} {126}},\ \bibinfo {pages} {080502} (\bibinfo {year}
  {2021})}\BibitemShut {NoStop}%
\bibitem [{\citenamefont {T\'oth}(2012)}]{PhysRevA.85.022322}%
  \BibitemOpen
  \bibfield  {author} {\bibinfo {author} {\bibfnamefont {G.}~\bibnamefont
  {T\'oth}},\ }\bibfield  {title} {\bibinfo {title} {Multipartite entanglement
  and high-precision metrology},\ }\href
  {https://doi.org/10.1103/PhysRevA.85.022322} {\bibfield  {journal} {\bibinfo
  {journal} {Phys. Rev. A}\ }\textbf {\bibinfo {volume} {85}},\ \bibinfo
  {pages} {022322} (\bibinfo {year} {2012})}\BibitemShut {NoStop}%
\bibitem [{\citenamefont {Imai}\ \emph {et~al.}(2024)\citenamefont {Imai},
  \citenamefont {Smerzi},\ and\ \citenamefont
  {Pezz{\`e}}}]{imai2024metrological}%
  \BibitemOpen
  \bibfield  {author} {\bibinfo {author} {\bibfnamefont {S.}~\bibnamefont
  {Imai}}, \bibinfo {author} {\bibfnamefont {A.}~\bibnamefont {Smerzi}},\ and\
  \bibinfo {author} {\bibfnamefont {L.}~\bibnamefont {Pezz{\`e}}},\ }\bibfield
  {title} {\bibinfo {title} {Metrological usefulness of entanglement and
  nonlinear hamiltonians},\ }\href@noop {} {\bibfield  {journal} {\bibinfo
  {journal} {arXiv preprint arXiv:2405.15703}\ } (\bibinfo {year}
  {2024})}\BibitemShut {NoStop}%
\bibitem [{\citenamefont {Juli\`a-Farr\'e}\ \emph {et~al.}(2020)\citenamefont
  {Juli\`a-Farr\'e}, \citenamefont {Salamon}, \citenamefont {Riera},
  \citenamefont {Bera},\ and\ \citenamefont
  {Lewenstein}}]{PhysRevResearch.2.023113}%
  \BibitemOpen
  \bibfield  {author} {\bibinfo {author} {\bibfnamefont {S.}~\bibnamefont
  {Juli\`a-Farr\'e}}, \bibinfo {author} {\bibfnamefont {T.}~\bibnamefont
  {Salamon}}, \bibinfo {author} {\bibfnamefont {A.}~\bibnamefont {Riera}},
  \bibinfo {author} {\bibfnamefont {M.~N.}\ \bibnamefont {Bera}},\ and\
  \bibinfo {author} {\bibfnamefont {M.}~\bibnamefont {Lewenstein}},\ }\bibfield
   {title} {\bibinfo {title} {Bounds on the capacity and power of quantum
  batteries},\ }\href {https://doi.org/10.1103/PhysRevResearch.2.023113}
  {\bibfield  {journal} {\bibinfo  {journal} {Phys. Rev. Res.}\ }\textbf
  {\bibinfo {volume} {2}},\ \bibinfo {pages} {023113} (\bibinfo {year}
  {2020})}\BibitemShut {NoStop}%
\bibitem [{\citenamefont {Rossini}\ \emph {et~al.}(2020)\citenamefont
  {Rossini}, \citenamefont {Andolina}, \citenamefont {Rosa}, \citenamefont
  {Carrega},\ and\ \citenamefont {Polini}}]{PhysRevLett.125.236402}%
  \BibitemOpen
  \bibfield  {author} {\bibinfo {author} {\bibfnamefont {D.}~\bibnamefont
  {Rossini}}, \bibinfo {author} {\bibfnamefont {G.~M.}\ \bibnamefont
  {Andolina}}, \bibinfo {author} {\bibfnamefont {D.}~\bibnamefont {Rosa}},
  \bibinfo {author} {\bibfnamefont {M.}~\bibnamefont {Carrega}},\ and\ \bibinfo
  {author} {\bibfnamefont {M.}~\bibnamefont {Polini}},\ }\bibfield  {title}
  {\bibinfo {title} {Quantum advantage in the charging process of
  sachdev-ye-kitaev batteries},\ }\href
  {https://doi.org/10.1103/PhysRevLett.125.236402} {\bibfield  {journal}
  {\bibinfo  {journal} {Phys. Rev. Lett.}\ }\textbf {\bibinfo {volume} {125}},\
  \bibinfo {pages} {236402} (\bibinfo {year} {2020})}\BibitemShut {NoStop}%
\bibitem [{\citenamefont {Rosa}\ \emph {et~al.}(2020)\citenamefont {Rosa},
  \citenamefont {Rossini}, \citenamefont {Andolina}, \citenamefont {Polini},\
  and\ \citenamefont {Carrega}}]{rosa2020ultra}%
  \BibitemOpen
  \bibfield  {author} {\bibinfo {author} {\bibfnamefont {D.}~\bibnamefont
  {Rosa}}, \bibinfo {author} {\bibfnamefont {D.}~\bibnamefont {Rossini}},
  \bibinfo {author} {\bibfnamefont {G.~M.}\ \bibnamefont {Andolina}}, \bibinfo
  {author} {\bibfnamefont {M.}~\bibnamefont {Polini}},\ and\ \bibinfo {author}
  {\bibfnamefont {M.}~\bibnamefont {Carrega}},\ }\bibfield  {title} {\bibinfo
  {title} {Ultra-stable charging of fast-scrambling syk quantum batteries},\
  }\href@noop {} {\bibfield  {journal} {\bibinfo  {journal} {Journal of High
  Energy Physics}\ }\textbf {\bibinfo {volume} {2020}},\ \bibinfo {pages} {1}
  (\bibinfo {year} {2020})}\BibitemShut {NoStop}%
\bibitem [{\citenamefont {Andolina}\ \emph {et~al.}(2024)\citenamefont
  {Andolina}, \citenamefont {Stanzione}, \citenamefont {Giovannetti},\ and\
  \citenamefont {Polini}}]{andolina2024genuine}%
  \BibitemOpen
  \bibfield  {author} {\bibinfo {author} {\bibfnamefont {G.~M.}\ \bibnamefont
  {Andolina}}, \bibinfo {author} {\bibfnamefont {V.}~\bibnamefont {Stanzione}},
  \bibinfo {author} {\bibfnamefont {V.}~\bibnamefont {Giovannetti}},\ and\
  \bibinfo {author} {\bibfnamefont {M.}~\bibnamefont {Polini}},\ }\bibfield
  {title} {\bibinfo {title} {Genuine quantum advantage in non-linear bosonic
  quantum batteries},\ }\href@noop {} {\bibfield  {journal} {\bibinfo
  {journal} {arXiv preprint arXiv:2409.08627}\ } (\bibinfo {year}
  {2024})}\BibitemShut {NoStop}%
\bibitem [{\citenamefont {Giovannetti}\ \emph {et~al.}(2003)\citenamefont
  {Giovannetti}, \citenamefont {Lloyd},\ and\ \citenamefont
  {Maccone}}]{giovannetti2003role}%
  \BibitemOpen
  \bibfield  {author} {\bibinfo {author} {\bibfnamefont {V.}~\bibnamefont
  {Giovannetti}}, \bibinfo {author} {\bibfnamefont {S.}~\bibnamefont {Lloyd}},\
  and\ \bibinfo {author} {\bibfnamefont {L.}~\bibnamefont {Maccone}},\
  }\bibfield  {title} {\bibinfo {title} {The role of entanglement in dynamical
  evolution},\ }\href@noop {} {\bibfield  {journal} {\bibinfo  {journal}
  {Europhysics Letters}\ }\textbf {\bibinfo {volume} {62}},\ \bibinfo {pages}
  {615} (\bibinfo {year} {2003})}\BibitemShut {NoStop}%
\bibitem [{\citenamefont {Xu}(2016)}]{xu2016detecting}%
  \BibitemOpen
  \bibfield  {author} {\bibinfo {author} {\bibfnamefont {Z.-Y.}\ \bibnamefont
  {Xu}},\ }\bibfield  {title} {\bibinfo {title} {Detecting quantum speedup in
  closed and open systems},\ }\href@noop {} {\bibfield  {journal} {\bibinfo
  {journal} {New Journal of Physics}\ }\textbf {\bibinfo {volume} {18}},\
  \bibinfo {pages} {073005} (\bibinfo {year} {2016})}\BibitemShut {NoStop}%
\bibitem [{Note1()}]{Note1}%
  \BibitemOpen
  \bibinfo {note} {The target state $|\psi (T)\rangle $ may include an
  arbitrary phase factor, i.e., $|\psi (T)\rangle =e^{i\phi }|\uparrow \rangle
  ^{\otimes N}$. However, in many cases, computations and proofs remain valid
  regardless of this phase choice. Therefore, we set it to 0 unless doing so
  would affect generality.}\BibitemShut {Stop}%
\bibitem [{\citenamefont {Deffner}\ and\ \citenamefont
  {Campbell}(2017)}]{deffner2017quantum}%
  \BibitemOpen
  \bibfield  {author} {\bibinfo {author} {\bibfnamefont {S.}~\bibnamefont
  {Deffner}}\ and\ \bibinfo {author} {\bibfnamefont {S.}~\bibnamefont
  {Campbell}},\ }\bibfield  {title} {\bibinfo {title} {Quantum speed limits:
  from heisenberg’s uncertainty principle to optimal quantum control},\
  }\href {https://doi.org/10.1088/1751-8121/aa86c6} {\bibfield  {journal}
  {\bibinfo  {journal} {Journal of Physics A: Mathematical and Theoretical}\
  }\textbf {\bibinfo {volume} {50}},\ \bibinfo {pages} {453001} (\bibinfo
  {year} {2017})}\BibitemShut {NoStop}%
\bibitem [{\citenamefont {Mandelstam}(1945)}]{mandelstam1945uncertainty}%
  \BibitemOpen
  \bibfield  {author} {\bibinfo {author} {\bibfnamefont {L.}~\bibnamefont
  {Mandelstam}},\ }\bibfield  {title} {\bibinfo {title} {The uncertainty
  relation between energy and time in nonrelativistic quantum mechanics},\
  }\href@noop {} {\bibfield  {journal} {\bibinfo  {journal} {J. Phys.(USSR)}\
  }\textbf {\bibinfo {volume} {9}},\ \bibinfo {pages} {249} (\bibinfo {year}
  {1945})}\BibitemShut {NoStop}%
\bibitem [{\citenamefont {Margolus}\ and\ \citenamefont
  {Levitin}(1998)}]{margolus1998maximum}%
  \BibitemOpen
  \bibfield  {author} {\bibinfo {author} {\bibfnamefont {N.}~\bibnamefont
  {Margolus}}\ and\ \bibinfo {author} {\bibfnamefont {L.~B.}\ \bibnamefont
  {Levitin}},\ }\bibfield  {title} {\bibinfo {title} {The maximum speed of
  dynamical evolution},\ }\href
  {https://doi.org/https://doi.org/10.1016/S0167-2789(98)00054-2} {\bibfield
  {journal} {\bibinfo  {journal} {Physica D: Nonlinear Phenomena}\ }\textbf
  {\bibinfo {volume} {120}},\ \bibinfo {pages} {188} (\bibinfo {year}
  {1998})}\BibitemShut {NoStop}%
\bibitem [{\citenamefont {Levitin}\ and\ \citenamefont
  {Toffoli}(2009)}]{PhysRevLett.103.160502}%
  \BibitemOpen
  \bibfield  {author} {\bibinfo {author} {\bibfnamefont {L.~B.}\ \bibnamefont
  {Levitin}}\ and\ \bibinfo {author} {\bibfnamefont {T.}~\bibnamefont
  {Toffoli}},\ }\bibfield  {title} {\bibinfo {title} {Fundamental limit on the
  rate of quantum dynamics: The unified bound is tight},\ }\href
  {https://doi.org/10.1103/PhysRevLett.103.160502} {\bibfield  {journal}
  {\bibinfo  {journal} {Phys. Rev. Lett.}\ }\textbf {\bibinfo {volume} {103}},\
  \bibinfo {pages} {160502} (\bibinfo {year} {2009})}\BibitemShut {NoStop}%
\bibitem [{\citenamefont {S\o{}rensen}\ and\ \citenamefont
  {M\o{}lmer}(2001)}]{PhysRevLett.86.4431}%
  \BibitemOpen
  \bibfield  {author} {\bibinfo {author} {\bibfnamefont {A.~S.}\ \bibnamefont
  {S\o{}rensen}}\ and\ \bibinfo {author} {\bibfnamefont {K.}~\bibnamefont
  {M\o{}lmer}},\ }\bibfield  {title} {\bibinfo {title} {Entanglement and
  extreme spin squeezing},\ }\href
  {https://doi.org/10.1103/PhysRevLett.86.4431} {\bibfield  {journal} {\bibinfo
   {journal} {Phys. Rev. Lett.}\ }\textbf {\bibinfo {volume} {86}},\ \bibinfo
  {pages} {4431} (\bibinfo {year} {2001})}\BibitemShut {NoStop}%
\bibitem [{\citenamefont {Szalay}(2019)}]{szalay2019k}%
  \BibitemOpen
  \bibfield  {author} {\bibinfo {author} {\bibfnamefont {S.}~\bibnamefont
  {Szalay}},\ }\bibfield  {title} {\bibinfo {title} {k-stretchability of
  entanglement, and the duality of k-separability and k-producibility},\ }\href
  {https://doi.org/https://doi.org/10.22331/q-2019-12-02-204} {\bibfield
  {journal} {\bibinfo  {journal} {Quantum}\ }\textbf {\bibinfo {volume} {3}},\
  \bibinfo {pages} {204} (\bibinfo {year} {2019})}\BibitemShut {NoStop}%
\bibitem [{SM()}]{SM}%
  \BibitemOpen
  \href@noop {} {}\bibinfo {note} {See Supplemental Material at
  URL-will-be-inserted-by-publisher for the proof of Theorem 1, the spin-$d/2$
  irreducible representation of the su(2) Lie algebra, the fully charging
  condition, and the numerical verification of the conjectured
  bound.}\BibitemShut {Stop}%
\bibitem [{\citenamefont {Yang}\ \emph {et~al.}(2024)\citenamefont {Yang},
  \citenamefont {Shi}, \citenamefont {Wan}, \citenamefont {Zhang},
  \citenamefont {Wang},\ and\ \citenamefont {Yang}}]{PhysRevA.109.012204}%
  \BibitemOpen
  \bibfield  {author} {\bibinfo {author} {\bibfnamefont {H.-Y.}\ \bibnamefont
  {Yang}}, \bibinfo {author} {\bibfnamefont {H.-L.}\ \bibnamefont {Shi}},
  \bibinfo {author} {\bibfnamefont {Q.-K.}\ \bibnamefont {Wan}}, \bibinfo
  {author} {\bibfnamefont {K.}~\bibnamefont {Zhang}}, \bibinfo {author}
  {\bibfnamefont {X.-H.}\ \bibnamefont {Wang}},\ and\ \bibinfo {author}
  {\bibfnamefont {W.-L.}\ \bibnamefont {Yang}},\ }\bibfield  {title} {\bibinfo
  {title} {Optimal energy storage in the tavis-cummings quantum battery},\
  }\href {https://doi.org/10.1103/PhysRevA.109.012204} {\bibfield  {journal}
  {\bibinfo  {journal} {Phys. Rev. A}\ }\textbf {\bibinfo {volume} {109}},\
  \bibinfo {pages} {012204} (\bibinfo {year} {2024})}\BibitemShut {NoStop}%
\bibitem [{\citenamefont {Yang}\ \emph {et~al.}(2025)\citenamefont {Yang},
  \citenamefont {Zhang}, \citenamefont {Wang},\ and\ \citenamefont
  {Shi}}]{PhysRevB.111.085410}%
  \BibitemOpen
  \bibfield  {author} {\bibinfo {author} {\bibfnamefont {H.-Y.}\ \bibnamefont
  {Yang}}, \bibinfo {author} {\bibfnamefont {K.}~\bibnamefont {Zhang}},
  \bibinfo {author} {\bibfnamefont {X.-H.}\ \bibnamefont {Wang}},\ and\
  \bibinfo {author} {\bibfnamefont {H.-L.}\ \bibnamefont {Shi}},\ }\bibfield
  {title} {\bibinfo {title} {Optimal energy storage and collective charging
  speedup in the central-spin quantum battery},\ }\href
  {https://doi.org/10.1103/PhysRevB.111.085410} {\bibfield  {journal} {\bibinfo
   {journal} {Phys. Rev. B}\ }\textbf {\bibinfo {volume} {111}},\ \bibinfo
  {pages} {085410} (\bibinfo {year} {2025})}\BibitemShut {NoStop}%
\bibitem [{\citenamefont {Gantmacher}\ and\ \citenamefont
  {Krein}(2002)}]{gantmacher2002oscillation}%
  \BibitemOpen
  \bibfield  {author} {\bibinfo {author} {\bibfnamefont {F.~R.}\ \bibnamefont
  {Gantmacher}}\ and\ \bibinfo {author} {\bibfnamefont {M.~G.}\ \bibnamefont
  {Krein}},\ }\href@noop {} {\emph {\bibinfo {title} {Oscillation matrices and
  kernels and small vibrations of mechanical systems: revised edition}}}\
  (\bibinfo  {publisher} {American Mathematical Society Rhode Island},\
  \bibinfo {year} {2002})\BibitemShut {NoStop}%
\bibitem [{\citenamefont {Karbach}\ and\ \citenamefont
  {Stolze}(2005)}]{PhysRevA.72.030301}%
  \BibitemOpen
  \bibfield  {author} {\bibinfo {author} {\bibfnamefont {P.}~\bibnamefont
  {Karbach}}\ and\ \bibinfo {author} {\bibfnamefont {J.}~\bibnamefont
  {Stolze}},\ }\bibfield  {title} {\bibinfo {title} {Spin chains as perfect
  quantum state mirrors},\ }\href {https://doi.org/10.1103/PhysRevA.72.030301}
  {\bibfield  {journal} {\bibinfo  {journal} {Phys. Rev. A}\ }\textbf {\bibinfo
  {volume} {72}},\ \bibinfo {pages} {030301} (\bibinfo {year}
  {2005})}\BibitemShut {NoStop}%
\end{thebibliography}%

\clearpage
\begin{widetext}
\textbf{Supplemental Material for ``Quantum Charging Advantage from Multipartite Entanglement''}
\maketitle

\section{Proof of Theorem 1}
\emph{Lemma 1.}-- If a pure state $\ket{\psi}$ of $N$ particles is not $(k+1)$-separable, then its entanglement depth satisfies ${\rm Ent}(\ket{\psi})\geq \lceil{N/k}\rceil$.

\emph{Proof:}  
Since a $k$-separable state is also $(k-1)$-separable, it suffices to show that for a $k$-separable but not $(k+1)$-separable state $\ket{\psi}$, the entanglement depth is lower-bounded as ${\rm Ent}(\ket{\psi})\geq \lceil{N/k}\rceil$.  
A $k$-separable state can be expressed as $\ket{\psi} = \bigotimes_{\ell=1}^{k} \ket{\psi_\ell}$, where $\ket{\psi_\ell}$ is a state of $N_\ell$ particles.  
Since $\ket{\psi}$ is not $(k+1)$-separable, each $\ket{\psi_\ell}$ must have an entanglement depth of ${\rm Ent}(\ket{\psi_\ell}) = N_\ell$.  
Thus, $\ket{\psi}$ is $(\max_\ell N_\ell)$-producible but not $(\max_\ell N_\ell - 1)$-producible, implying that ${\rm Ent}(\ket{\psi}) = \max_\ell N_\ell$.  
Given that $\sum_{\ell=1}^{k} N_\ell = N$, it follows that $\max_\ell N_\ell \geq \lceil{N/k}\rceil$, which establishes the result.

\emph{Theorem 1.}-- 
Let $p_0$ and $p_{\bar 0}$ be the superposition coefficients of a locally orthonormal pair in an arbitrary  $N$-qubit pure state, i.e., $\ket{\psi}=p_0\ket{v_0}+p_{\bar 0}\ket{v_{\bar 0}}+\cdots$.
The entanglement depth of $\ket{\psi}$ is  then bounded from below as
\begin{eqnarray}\label{Detect-Ent-1}
{\rm Ent}(\ket{\psi})\!\geq\!\left\lceil\frac{N}{\left\lfloor\log_21/|p_0p_{\bar 0}|\right\rfloor}\right\rceil.
\end{eqnarray}

\emph{Proof.}--
Since $\ket{v_0}$ and $\ket{v_{\bar 0}}$ are locally orthogonal, a local unitary transformation can always be applied to transform $\ket{\psi}$ into $\ket{\psi'}\!=\! p_0 \ket{\up}^{\otimes N } + p_{\bar 0} \ket{\dw}^{\otimes N } + \cdots$, without changing its entanglement depth.
Suppose that $\ket{\psi'}$ can be expressed as a $k$-separable state, $\ket{\psi'}\!=\!\otimes_{\ell=1}^k\ket{\psi_\ell'}$, where  
\begin{eqnarray}
\ket{\psi_\ell'}=a_\ell\ket{\underbrace{\up\up\cdots\up}_{ N_\ell}}+b_\ell \ket{\underbrace{\dw\dw\cdots\dw}_{ N_\ell}}+\cdots.
\end{eqnarray} 
The normalization condition $\ex{\psi_\ell'|\psi_\ell'}\!=\!1$ implies that  $|a_\ell|^2+|b_\ell|^2\leq 1$.
Given that $\prod_{\ell=1}^k a_\ell\!=\! p_0$ and $\prod_{\ell=1}^k b_\ell\!=\! p_{\bar 0}$, we deduce   
\begin{align}
1\geq \prod_{\ell=1}^k (|a_\ell|^2+|b_\ell|^2)\geq 2^k\prod_{\ell=1}^k |a_\ell||b_\ell|=2^k |p_0p_{\bar 0}|,
\end{align}
which provides the bound 
$k\!\leq\! \log_2 (1/|p_0p_{\bar 0}|)$.
Finally, by applying Lemma 1,  inequality~(\ref{Detect-Ent-1}) is satisfied for $\ket{\psi'}$.
Since ${\rm Ent}(\ket{\psi}) \!=\! {\rm Ent}(\ket{\psi'})$, the proof is complete.

\section{Irreducible Representation of the SU(2) Lie Algebra}

The spin-$d/2$ irreducible representation of the SU(2) algebra generators is given by
\begin{eqnarray}
J^x &=& \frac{1}{2}
\begin{pmatrix}
0 & f_1 & & \\
f_1 & 0 & f_2 & \\
& \ddots & \ddots & \ddots & \\
& & f_{d-1} & 0 & f_d \\
& & & f_d & 0
\end{pmatrix}, \quad
J^y = \frac{1}{2i}
\begin{pmatrix}
0 & -f_1 & & \\
f_1 & 0 & -f_2 & \\
& \ddots & \ddots & \ddots & \\
& & f_{d-1} & 0 & -f_d \\
& & & f_d & 0
\end{pmatrix}, \quad
J^z =
\begin{pmatrix}
-\frac{d}{2} & 0 & & \\
0 & -\frac{d}{2} + 1 & 0 & \\
& \ddots & \ddots & \ddots & \\
& & 0 & \frac{d}{2} - 1 & 0 \\
& & & 0 & \frac{d}{2}
\end{pmatrix},
\end{eqnarray}
where $f_k \equiv \sqrt{k(d-k+1)}$.

\section{Condition for Fully Charging }

\subsection{SU(2) charging schemes}
We analyze fully charging schemes governed by the unitary evolution  $\ket{\psi(t)} = e^{-iH_dt} \ket{\psi(0)}$, where  
\begin{equation}
H_d = \alpha_1 J^x + \alpha_2 J^y + \alpha_3 J^z + \frac{d}{2} |\vec{\alpha}|,
\end{equation}
and the initial state is $\ket{\psi(0)} = \ket{\downarrow}^{\otimes N}$.  
The effective Hamiltonian $H_d$ acts on a subspace $\mathcal H_{d+1} \subset \mathcal H_2^{\otimes N}$ spanned by the initial state $\ket{u_0} \equiv \ket{\downarrow}^{\otimes N}$, the target state $\ket{u_d} \equiv \ket{\uparrow}^{\otimes N}$, and the intermediate states $\ket{u_1}, \dots, \ket{u_{d-1}}$. 
The charging condition  $\ket{\psi(T)} = \ket{\uparrow}^{\otimes N}$ is equivalent to  
\begin{equation}\label{fully-charging-0}
\langle \psi(0) | J^z(T) | \psi(0) \rangle = \frac{d}{2},
\end{equation}
where $J^z(T) = e^{iH_dT} J^z e^{-iH_dT}$.
The condition~(\ref{fully-charging-0}) is also equivalent to  $J^z(T) = -J^z$. 
From Lie algebra theory, the necessary and sufficient condition for this to hold is $\alpha_3 = 0$.

\subsection{Tridiagonal charging schemes}

For generalized charging schemes, we consider the tridiagonal Hamiltonian $H_{t,d+1}$
 \begin{equation}
    \label{eq:tridiagonal_H}
    H_{t,d+1}\left(\{b_j\}_{j=1}^d\right) = \begin{pmatrix}
        0 & b_1 & 0 & \cdots & 0 & 0 & 0\\
        b_1 & 0 & b_2 & \cdots & 0 & 0 & 0 \\
        0 & b_2 & 0 & \cdots & 0 & 0 & 0 \\
        \vdots & \vdots & \vdots & \ddots & \vdots & \vdots & \vdots \\
        0 & 0 & 0 & \cdots & 0 & b_{d-1} & 0 \\
        0 & 0 & 0 & \cdots & b_{d-1} & 0 & b_{d} \\
        0 & 0 & 0 & \cdots & 0 & b_{d} & 0 \\
    \end{pmatrix},\quad b_j>0,
\end{equation}
acting on subspace $\mathcal H_{d+1} \subset \mathcal H_2^{\otimes N}$ spanned by $\{|u_0\rangle,|u_1\rangle,\cdots, |u_d\rangle\}$.
The condition for fully charging in tridiagonal charging models is
    \begin{equation}
    \label{eq:optimal_charing_condition}
        e^{-iH_{t,d+1}T}|u_0\rangle = e^{i\phi_0}|u_d\rangle.
    \end{equation}

For eigenvalues $E_k$ and eigenvectors $\{|E_k\rangle\}_{k=0}^d$ of $H_{t,d+1}$ (ordered by $E_0<E_1<\cdots<E_d$), the mirror symmetry
    \begin{equation}
    \label{eq:mirror_sym}
        b_k = b_{d+1-k},
    \end{equation}
implies~\cite{gantmacher2002oscillation} 
    \begin{equation}\label{mirror-symmetry}
        \langle u_0|E_k\rangle = (-1)^k\langle u_{d}|E_k\rangle.
    \end{equation}
Combined with the energy spectrum condition~\cite{PhysRevA.72.030301}
    \begin{equation}\label{fully-condition-1}
        E_k T = (2m_k-k)\pi + \phi_0, \ m_k\in \mathbb Z,
    \end{equation}
the fully charging dynamics can be realized since
\begin{equation}
        e^{-iH_{t,d+1}\tau}|u_0\rangle = \sum_{k=0}^d e^{-iE_kT}|E_k\rangle\langle E_k|u_0\rangle \overset{Eq.~(\ref{mirror-symmetry})}{=} \sum_{k=0}^d (-1)^k e^{-iE_kT}|E_k\rangle\langle E_k|u_d\rangle
\overset{Eq.~(\ref{fully-condition-1})}{=}
e^{i\phi}\ket{u_d}.
    \end{equation}

\subsection{Example: \emph{d}=3 tridiagonal charging Hamiltonian}
For $d=3$, the mirror-symmetric tridiagonal Hamiltonian is
\begin{equation}
    H_{t,4} = \begin{pmatrix}
        0 & b_1 & 0 & 0 \\
        b_1 & 0 & b_2 & 0 \\
        0 & b_2 & 0 & b_1 \\
        0 & 0 & b_1 & 0 \\
    \end{pmatrix}.
\end{equation}
Eigenanalysis yields parameters $b_1 = \sqrt{\lambda_1 \lambda_2}$ and  $b_2 = \lambda_1-\lambda_2$, with eigenvalues  $    -\lambda_1 \!<\!-\lambda_2 \!<\! \lambda_2 \!<\! \lambda_1$.
The charging dynamics is obtained as 
\begin{equation}\label{evloved-state}
\ket{\psi(t)}=e^{-iH_{t,4} t}|u_0\rangle = \frac{1}{\lambda_1+\lambda_2}\begin{pmatrix}
        \lambda_2\cos\left(\lambda_1 t\right) + \lambda_1\cos\left(\lambda_2 t\right) \\
        -i\sqrt{\lambda_1\lambda_2}\left(\sin\left(\lambda_1 t\right)+\sin\left(\lambda_2 t\right)\right) \\
        \sqrt{\lambda_1\lambda_2}\left(\cos\left(\lambda_1 t\right)-\cos\left(\lambda_2 t\right)\right) \\
        i\left(\lambda_1\sin\left(\lambda_2 t\right)-\lambda_2\sin\left(\lambda_1 t\right)\right) \\
    \end{pmatrix}\equiv \sum_{j=0}^3 p_j(t)\ket{u_j}
\end{equation}
The time $T$ required for full charging is thus determined by 
\begin{eqnarray}\label{eq0}
   \frac{\abs{\lambda_1\sin\left(\lambda_2 T\right)-\lambda_2\sin\left(\lambda_1 T\right)}}{\lambda_1+\lambda_2}=1. 
\end{eqnarray}
To ensure the existence of a solution of $T$, we require either (i) $\sin(\lambda_2T)\!=\!-\sin(\lambda_1T)=1$,
or
(ii) $\sin(\lambda_1T)\!=\!-\sin(\lambda_2T)=1$.
Denoting $k=\lambda_1/\lambda_2>1$,  case (i) reduces to 
\begin{eqnarray}\label{time-T}
    \frac{2T}{\pi}=\frac{1+4m}{\lambda_2}=\frac{4n-1}{\lambda_1}, \quad m,n\in\mathbb Z.
\end{eqnarray}
Thus,  the possible values of $k$ in case (i) are 
\begin{eqnarray}\label{case1}
     k=\frac{4n-1}{4m+1}>1,\quad m,n\in\mathbb Z.
 \end{eqnarray} 
Similarly, in case (ii) the possible values of $k$ are 
\begin{eqnarray}\label{case2}
     k=\frac{4m+1}{4n-1}>1,\quad m,n\in\mathbb Z.
 \end{eqnarray}

\begin{figure}[t]
      \includegraphics[width=0.5\columnwidth]{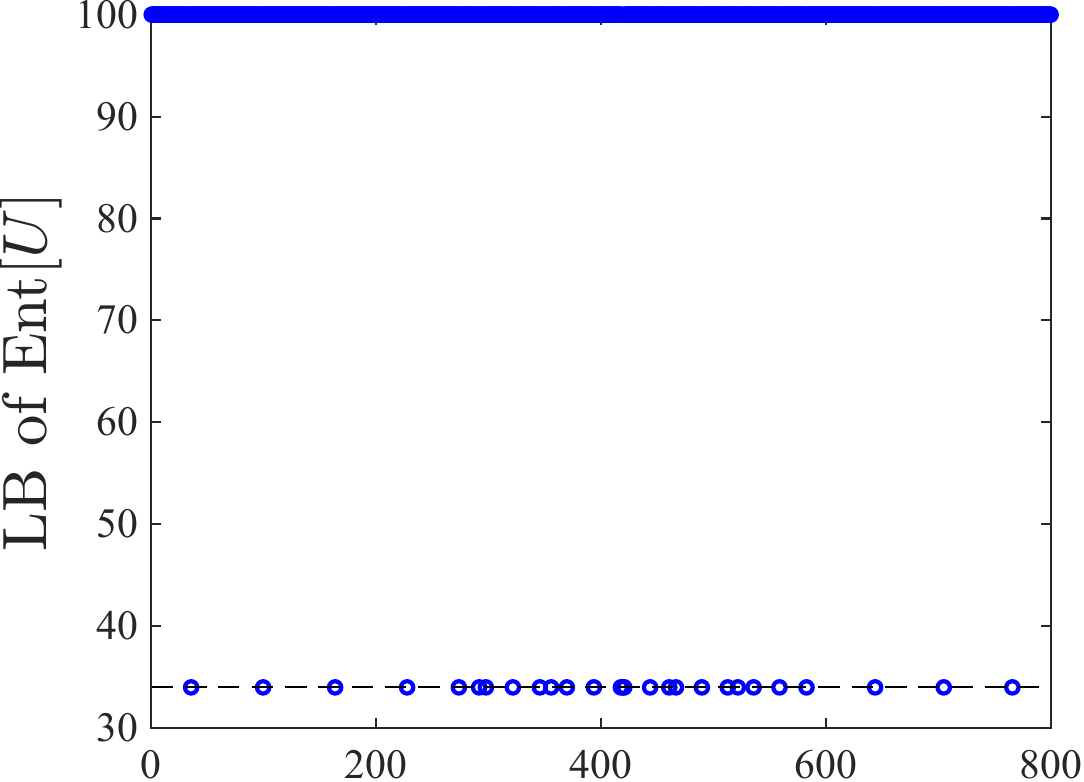}
      \caption{
Verification of conjecture~(\ref{conjucture-1}). 
Blue circles denote the maximum lower bound (LB) of entanglement depth, Eq.~(\ref{appendix-value}), in different fully charging schemes.
The fully charging schemes are generated by considering the fully charging conditions~(\ref{case1}) and (\ref{case2}).
The charging schemes lying in the black dashed line correspond to the SU(2) charging schemes.
However, blue circles with a value of 100 correspond to the tridiagonal charging schemes without SU(2) symmetry.
Here $N\!=\!100$ qubits are considered.
}\label{fig}
\end{figure}

We note that the eigenspectrum for SU(2) charging is linear, corresponding to the $k=3$ case.
According to Eq.~(\ref{evloved-state}) by substituting 
\begin{eqnarray}\label{THM1-use}
    |p_0(t)p_3(t)|&=&\frac{1}{(\lambda_1+\lambda_2)^2}\abs{[\lambda_2\cos(\lambda_1t)+\lambda_1\cos(\lambda_2t)][\lambda_1\sin(\lambda_2t)-\lambda_2\sin(\lambda_1t)]},
\end{eqnarray}
into the lower bound of Theorem 1 and then maximizing it over time $t$, we obtain the value 
\begin{align}\label{appendix-value}
  \max_t\left\lceil\frac{N}{\left\lfloor\log_21/|p_0(t)p_{3}(t)|\right\rfloor}\right\rceil,
\end{align}
and plot it in
 Fig.~\ref{fig} under the fully charging conditions~(\ref{case1}) and (\ref{case2}).
Theorem~1 implies that 
\begin{align}\label{appexdix-bound}
 {\rm Ent}[U]\!\geq\! \max_t\left\lceil\frac{N}{\left\lfloor\log_21/|p_0(t)p_{3}(t)|\right\rfloor}\right\rceil.
\end{align}
The blue circles in Fig.~\ref{fig} 
, marked with the black dashed line, correspond to the SU(2) case with $k=3$.
For these charging schemes, as discussed in the main text, the conjectured bound is satisfied.
For other charging schemes (blue circles with value 100), the bound~(\ref{appexdix-bound}) implies ${\rm Ent}[U]\!=\!N$, which observably satisfies the conjectured bound  ${\rm Ent}[U] \geq \left\lceil N\eta^2
\right\rceil$ for any $\eta$.

\end{widetext}

\end{document}